\documentclass{article}
\usepackage[margin=2cm]{geometry}
\usepackage{graphicx,bm,amsmath,amssymb,latexsym,psfrag,epsfig,euscript, multirow,color,verbatim,cancel}
\usepackage{mathrsfs}
\usepackage{float}

\allowdisplaybreaks
\begin{document}
\newcommand{\PL}{\mathcal{P}_{\mathrm L}}
\newcommand{\PR}{\mathcal{P}_{\mathrm R}}
\newcommand{\gs}{g_{\mathrm s}}
\newcommand{\be}{\begin{equation}}
\newcommand{\ee}{\end{equation}}
\newcommand{\beq}{\begin{equation}}
\newcommand{\eeq}{\end{equation}}
\newcommand{\bean}{\begin{eqnarray*}}
\newcommand{\eean}{\end{eqnarray*}}
\newcommand{\bea}{\begin{eqnarray}}
\newcommand{\eea}{\end{eqnarray}}
\newcommand{\MPl}{M_{\mathrm{Pl}}}
\newcommand{\Td}{T_{\mathrm{d}}}
\newcommand{\ptl}{\partial}
\newcommand{\benum}{\begin{enumerate}}
\newcommand{\eenum}{\end{enumerate}}
\newcommand{\bi}{\begin{itemize}}
\newcommand{\ei}{\end{itemize}}
\newcommand{\tb}{\tilde{B}}
\newcommand{\tw}{\tilde{W}}
\newcommand{\tg}{\tilde{g}}

\begin{titlepage}

\begin{flushright}
\end{flushright}

\vspace{0.2cm}
\begin{center}
\Large\bf
Natural Baryogenesis from Unnatural Supersymmetry
\end{center}

\vspace{0.2cm}
\begin{center}
{\sc Yanou Cui\footnote{E-mail:cuiyo@umd.edu~~~~~~~~~~~~~~~~~~~~~~~~{UMD-PP-013-012 }}
 }\\
\vspace{0.4cm}
{\sl Department of Physics, University of Maryland, College Park, MD 20742, USA.}
\end{center}
%\today
\vspace{0.2cm}
\begin{abstract}\vspace{0.2cm}
\noindent
We demonstrate here that the mini-split version of the \textit{Minimal} Supersymmetric Standard Model (MSSM) including R-parity violating couplings naturally provides all the necessary ingredients for a novel baryogenesis mechanism. The baryogenesis is triggered by the late decay of a TeV scale bino after its thermal freezeout. A $\mu$-term larger than the sfermion masses is necessary for obtaining sufficient baryon asymmetry. Two example models of direct baryogenesis and leptogenesis are proposed, with viable parameter spaces presented. The cosmological conditions for the models--in particular, the requirements of a long lifetime of bino and sufficient baryon asymmetry--point towards the mini-split scale of $\sim$100-1000 TeV for the sfermion masses. This provides an independent motivation for mini-split SUSY, along with the constraints from flavor physics and Higgs mass measurement. We also discuss the potential multi-pronged search for the signatures of such models, including those at the Large Hadron Collider (LHC) and the low energy experiments at the intensity frontier. 
\end{abstract}
\vfil

\end{titlepage}

\tableofcontents

\section{Introduction}
\indent

    Despite our familiarity with baryonic matter, the origin of the cosmic abundance of baryons $\Omega_{\Delta B}\simeq4\%$ remains one of the most prominent questions unanswered by the Standard Model (SM) of particle physics, just as puzzling as the questions around dark matter (DM). Meanwhile, the SM itself faces the Planck-electroweak (EW) hierarchy problem, suggesting new physics emerging around the weak scale, such as supersymmetry (SUSY). Most of the existing baryogenesis mechanisms introduce new particles and interactions at a much higher mass scale, or introduce weak scale particles unrelated to the hierarchy problem. It would be economic and desirable to have a weak scale baryogenesis mechanism enabled by new physics already motivated by solving the hierarchy problem. This is in similar spirit to the conventional preference for the SUSY LSP as a WIMP DM candidate. In this work, we explore the possibility and viability of such a baryogenesis mechanism in the framework of the \textit{Minimal} Supersymmetric Standard Model (MSSM) with mini-split spectrum\cite{ArkaniHamed:2004fb,splitsusy,Arvanitaki:2012ps,ArkaniHamed:2012gw}, a SUSY scenario that has drawn rising interest lately in light of the recent constraints from the LHC data as well as existing flavor physics bounds.\\

    In the conventional MSSM, the superpartners of SM particles all have weak scale masses, in order to stabilize the Higgs mass without fine-tuning. However, SUSY naturalness has been challenged by the recent collider search limits on superpartner masses as well as the measured Higgs mass around 125 GeV\cite{susytension}. Various interesting possibilities still exist to accommodate naturalness in SUSY, such as those involving light stops, R-parity violation (RPV) or contents beyond the MSSM (see \cite{naturalsusy} for a partial list). Nonetheless, many of these options require elaborate structure or hidden sources of tuning. After all, despite its appeals, naturalness is not a fundamental principle. If we opt to give up full naturalness, the ``mini-split'' version of the MSSM is an alternative simple scenario that is comfortably consistent with current constraints from the LHC as well as low energy experiments related to flavor physics and CP violation. In this version of SUSY, gaugino masses can still be $\sim \rm TeV$, while scalar superpartners have masses $\sim10^2-10^3\rm ~TeV$. Such spectrum is ubiquitous in existing SUSY breaking models such as in anomaly mediation without sequestering. Split MSSM also preserves major merits of conventional MSSM, such as improving gauge coupling unification. Notice that despite the moderate hierarchy, mini-split SUSY still accomplishes a significant improvement in naturalness compared to the Standard Model alone which suffers from the severe Planck-EW hierarchy problem. It is therefore plausible that the imperfect realization of naturalness in mini-split SUSY may result from a compromise between naturalness and some other principle, such as vacuum/environmental selection, which has been discussed in earlier literature\cite{ArkaniHamed:2004fb, ArkaniHamed:2005yv, Giudice:2006sn}. The results of this work may provide an example for such an interpretation, by demonstrating the natural viability for baryogenesis with the contents that are intrinsic to the minimal model of mini-split SUSY with RPV. \\
    
      Weak scale SUSY with conserved R-parity is well-known for its natural interface with modern cosmology: it provides neutralino LSP as a WIMP DM candidate. Recently, R-parity violating (RPV) SUSY has drawn more attention for the purpose of saving naturalness in SUSY, although RPV interactions may well exist in the context of high scale SUSY or split SUSY. We will demonstrate that the ``unnatural'' split SUSY, together with RPV, may ``naturally'' resolve the prominent puzzle about the origin of baryon asymmetry. By quick assessment, one can see the natural plausibility to satisfy all the Sakharov conditions \cite{Sakharov:1967dj} for baryogenesis within the split MSSM. Firstly, RPV MSSM generically comes with new sources of CP violation and B-( L-) number violation, which are necessary for baryogenesis. If all the superpartners are at weak scale as in the conventional MSSM, with generic flavor structure, sizable $\cancel{\rm CP}$ or RPV is strongly constrained by current experiments \cite{Gabbiani:1996hi,Martin:1997ns}. In contrast, if scalar superpartners take heavy masses $m_0\gtrsim100-1000$ TeV as in the split SUSY, the relevant bounds can be evaded in general, which makes it much safer to exploit large $\cancel{\rm CP}$ and RPV couplings for baryogenesis. Meanwhile, the large mass hierarchy between gauginos and scalar superpartners can result in a natural long lifetime of a gaugino $\chi$ with 3-body RPV decay such as $\chi\rightarrow udd$, suppressed by heavy $m_0$. The out-of-equilibrium late decay of $\chi$ provides the third Sakharov condition. \\
      
      Motivated by these observations, we investigate in detail the realistic viability of a baryogenesis mechanism within the framework of split MSSM. As we will show in the later sections, despite the apparent plausibility, it is subtle and challenging to get sizable CP asymmetry and consequently sufficient baryon asymmetry within this simply minimal setup. Nonetheless, an interesting scenario is found to be successful where weak scale leptogenesis or direct baryogenesis is triggered by a TeV scale bino when it decays after its thermal annihilation freezes out of equilibrium. Interestingly, considering the relevant cosmological constraints such baryogenesis models independently favor the ``mini-split'' scale of $m_0\sim 100-1000$ TeV. A wino or gluino that is lighter than bino, together with a $\mu$-term larger than $m_0$ are key to generate enough $\Omega_{\Delta B}$.\\
      
       In regards to UV completion, wino LSP is a generic prediction from anomaly mediated SUSY breaking which also generically generates a split spectrum\cite{Randall:1998uk}. The possibility of gluino LSP has also been well studied\cite{Raby:1997bpa, Baer:1998pg}. Regarding $\mu$-term, as the only supersymmetric parameter in the MSSM, in principle, it can be at a high scale well above weak scale, and different from both $m_{\rm gaugino}$ and $m_0$. Motivated by full naturalness principle and well-tempered dark matter candidates in R-parity conserving SUSY, the conventionally well studied region is $\mu\sim m_{\rm gaugino}\sim m_{\rm EW}$. Obtaining such a small weak scale $\mu$ term in fact often requires elaborate structure, and is entitled as ``solving the $\mu$-problem''. The more generic case of heavy higgsino, hence large $\mu\sim m_0\gg m_{\rm gaugino}$ has been explored in \cite{Cheung:2005ba}, and more recently in \cite{Arvanitaki:2012ps,ArkaniHamed:2012gw}. The spectrum preferred by our baryogenesis mechanism, $\mu\gg m_0$, is as plausible as those that have been extensively considered. As recently pointed out in \cite{ArkaniHamed:2012gw}, despite the associated tuning, this region is an intriguing case and innocuous in terms of phenomenology. Our results for baryogenesis may motivate more consideration for the UV explanation and implications of this parameter region.\\

       We want to comment that the general idea of baryogenesis from late decay of a metastable WIMP after its thermal freezeout was proposed in our earlier work\cite{Cui:2012jh}, where we gave model example in the context of natural SUSY with RPV. The baryogenesis models studied in this work can be seen as an embedding of that general idea in the scenario of split SUSY. As discussed in \cite{Cui:2012jh} as well as in earlier literature\cite{Barbier:2004ez}, RPV interactions in SUSY models tend to erase any primordial baryon asymmetry generated by conventional baryogenesis at high scale. The baryogenesis via RPV decay of a metastable WIMP provides a natural remedy to this potential problem, by resetting the asymmetry at a later time.     \\
   
    The rest of the paper is outlined as follows. In Section.\ref{sec:model} we first briefly review the general paradigm proposed in \cite{Cui:2012jh}, then discuss the necessary contents and spectrum for baryogenesis in split MSSM and compute the CP asymmetry for two example models. In Section.\ref{sec:results} we compute the baryon abundance and demonstrate examples of viable parameter space considering cosmological constraints. Section.\ref{sec:pheno} includes discussion about implications on phenomenology and possible experimental signatures. Finally we conclude with outlooks in Section.\ref{sec:concl}.
    
\section{The Model}\label{sec:model}
\subsection{Baryogenesis from Metastable WIMP decay: A Review}
   We first briefly review the general setup of metastable WIMP triggered baryogenesis as proposed in \cite{Cui:2012jh}. The conventional ``WIMP'' refers to a stable weak scale particle with weak interactions with the SM fields, which can be a good dark matter candidate according to the ``WIMP miracle''. Nonetheless, just like the known particles in the SM, in general a WIMP can have diverse lifetimes, depending on symmetry protection, mediator mass and couplings associated with possible decay channels. In particular, a metastable WIMP $\chi$ may experience a thermal freezeout stage just like a WIMP dark matter, but then later decays. Such decay can trigger baryogenesis if the process involves CP and B-(L-) violations. The freezeout of $\chi$ occurs when $\Gamma_{\rm ann}\sim H$, where $\Gamma_{\rm ann}$ is the annihilation rate, $H$ is the Hubble expansion rate. Suppose there is no later decay of $\chi$ (i.e. the $\chi$ lifetime $\tau\rightarrow\infty$), as if it were a DM candidate, its ``would-be'' relic abundance is set by its annihilation cross section:
      \bea
   \nonumber
   \Omega_{\chi}^{\tau\rightarrow\infty}&\simeq0.1&\frac{\alpha_{\rm weak}^2/(\rm TeV)^2}{\langle\sigma_{\rm A} v\rangle}\\
   &\simeq&0.1\left(\frac{g_{\rm weak}}{g_\chi}\right)^4\left(\frac{m_{\rm med}^4}{m_\chi^2\cdot\rm TeV^2 }\right)
\label{wimpmiracle}.
    \eea
    Apparently, with a weak scale ${\langle\sigma_{\rm A} v\rangle}$, $\Omega^{\tau\rightarrow\infty}_\chi$ is in the right ballpark of the observed $\Omega_{DM}\approx23\%$, which is the well-known ``WIMP miracle'' for dark matter. The second line in eq.(\ref{wimpmiracle}) manifests the dependence on model parameters in the generic case of a heavier scalar mediator. We assume that at a later stage after the thermal freezeout, $\chi$ decays in a $\cancel{CP}$ and $\cancel{B}(\cancel{L})$ way, which allows us to simply treat thermal freezeout and decay as decoupled processes. The resultant baryon abundance is directly proportional to the abundance of $\chi$ at the end of its freezeout, or its ``would-be'' relic abundance $\Omega^{\tau\rightarrow\infty}_\chi$ as given in eq.(\ref{wimpmiracle}). The assumption that $\chi$ decay after its thermal annihilation freezes out of equilibrium and after the thermal bath temperature falls below $\chi$ mass automatically suppress $\cancel{B}$($\cancel{L}$) washout processes such as inverse decay. With negligible washout effect, we obtain: 
    \beq
   \Omega_{\Delta B}=\epsilon_{\rm CP}\frac{m_p}{m_{\chi_B}}\Omega_{\chi_B}^{\tau\rightarrow\infty}\cdot\zeta,
    \label{omegaB}
    \eeq
     where $\epsilon_{CP}$ is the CP asymmetry of the decay, typically suppressed by 1-loop factor or more if there is a heavy mediator, while $\frac{m_p}{m_\chi}\sim 10^{-3}-10^{-2}$ for $\chi$ of weak scale mass. The addition factor $\zeta=1$ for direct baryogenesis occurring after electroweak phase transition. If $\chi$ decays before the phase transition, $\zeta$ represents the sphaleron distribution factor: $\zeta=51/79$ for direct baryogenesis, $\zeta=28/79$ for leptogenesis\cite{Chen:2007fv}.\\
     
     Now considering that $\Omega_{\Delta B}\approx\frac{1}{5}\frac{m_p}{m_\chi}\Omega_{DM}$ and $\epsilon_{CP}\frac{m_p}{m_\chi}\lesssim10^{-4}$, we find $\Omega^{\tau\rightarrow\infty}_\chi\gg\Omega_{DM}$ is necessary based on eqs.(\ref{wimpmiracle},\ref{omegaB}). This tells us that the viable baryon parent should be a rather weakly coupled WIMP (compared to a dark matter WIMP), with large ``overabundance'' in the limit where it is stable and may be considered as a DM candidate. Notice that if the suppression only comes from 1-loop factor, the required overabundance can be achieved by $O(1)$ adjustment of couplings or mediator mass associated with the baryon parent WIMP, as we can see from the second line in eq.(\ref{wimpmiracle}) which manifests the power law dependence on model parameters in a generic example. \\
     
\begin{figure}
   \begin{center}
        \includegraphics[height=80mm]{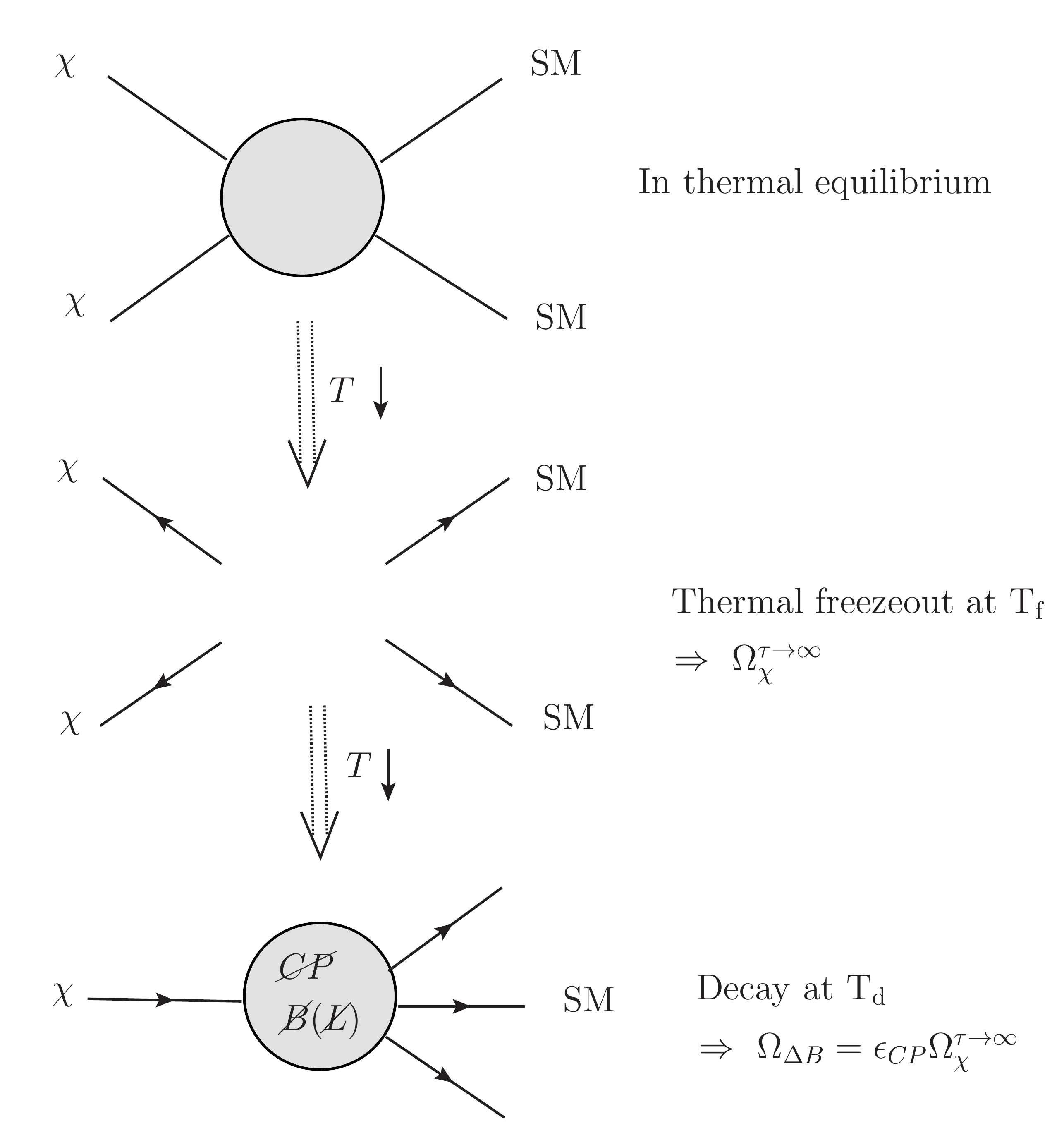} 
     \end{center} 
        \caption{Illustration of the cosmic evolution of WIMP baryon parent $\chi$. Dashed double arrow indicates the arrow of time, along which the temperature drops.}  
        \label{cartoon}   
\end{figure}

We illustrate the key physics processes for this novel baryogenesis mechanism in Fig.(\ref{cartoon}).
         
    \subsection{Bino as the baryon parent: $\tilde{B}\rightarrow \Delta{B}$}
    \indent
    Now we discuss how the split MSSM can fit into the above general paradigm for low scale baryogenesis. As discussed in the Introduction, in the split MSSM with RPV a weak scale fermionic superpartner can naturally have a long life-time due to the much heavier sfermion mediator in its RPV decay. The requirement of $\cancel{\rm CP}$ and $\cancel{B}$($\cancel{L}$) in the decay selects Majorana gauginos (rather than Dirac Higgsinos) as candidates for the WIMP $\chi$ as baryon parent. Furthermore, as reviewed earlier in this section, a would-be overabundance of $\chi$ is necessary for obtaining sufficient $\Omega_{\Delta B}$ as observed. Obviously, within the MSSM bino is the only viable candidate satisfying all the above conditions. As we will discuss in detail later, the major annihilation channel in pure bino limit is $\tb\tb\rightarrow HH$ with cross section ${\langle\sigma_{\rm A} v(\tilde{B})\rangle}\propto\mu^{-2}$.  Therefore $\mu\gg m_{W}$ is crucial to generate a large ``overabundance'' to compensate $\epsilon_{CP}\frac{m_p}{m_\chi}$.\\
       
      Now we write down the MSSM Lagrangian terms that are potentially relevant for the baryogenesis mechanism considered here. Although not a viable baryon parent, wino or gluino is crucial for generating CP asymmetry in bino decay via interference diagram as we will discuss in Section.\ref{secsec: CPV}. So we include the terms involving them as well. We also include both $\cancel{L}$ and $\cancel{B}$ trilinear RPV interactions, although in order to satisfy bound on proton lifetime either $\cancel{L}$ or $\cancel{B}$ needs to be absent or suppressed in a realistic model. The relevant Lagrangian terms are as follows: 
      \bea
      W&=&\mu H_uH_d+\frac{1}{2}\lambda^{ijk}L_iL_j\bar{e}_k+\lambda'^{ijk}L_iQ_j\bar{d}_k+\frac{1}{2}\lambda^{''ijk}\bar{u}_i\bar{d}_j\bar{d}_k+h.c.\\
      \mathcal{L}_{\rm gauge}&=&\frac{\sqrt{2}}{2}g_1(H_u^*\tilde{H}_u\tilde{B}-H_d^*\tilde{H}_d\tilde{B})+\sqrt{2}g_1Y_{{f}_{L/R,i}}\tilde{f}_i^{*L/R,\alpha}{f}_i^{L/R,\alpha}\tilde{B}\\\nonumber
&+&\sqrt{2}g_2\tilde{f}_i^{*L/R,\alpha}T^a{f}_i^{L/R,\alpha}\tilde{W}^a+\sqrt{2}g_3\tilde{f}_i^{*L/R,\alpha}T^a{f}_i^{L/R,\alpha}\tilde{g}^a+h.c.     \\
\mathcal{L}_{\rm soft}&=&-\frac{1}{2}M_1\tilde{B}\tilde{B}-\frac{1}{2}M_2\tilde{W}\tilde{W}-\frac{1}{2}M_3\tilde{g}\tilde{g}-\tilde{f}_i^{*L/R,\alpha}{\bf{(m_{L/R,\alpha}^2)}}_{ij}\tilde{f}_j^{L/R,\alpha}+h.c.,
       \eea
where $i, j$ are family indices, $\alpha$ labels a species with certain gauge charges (such as $u, L$ type), $L/R$ indicates left-handed or right-handed. CP violation can come from complex phases in $M_i$ and $\bf{(m_{L/R,\alpha}^2)}_{ij}$. Notice that although gaugino interactions are originally flavor diagonal in gauge basis, in case of split SUSY, flavor mixings from sfermion mass matrix $\bf{(m_{L/R,\alpha}^2)}_{ij}$ can generally be $O(1)$ with large CP phases, without violating experimental constraints. Therefore here the flavor violating gaugino couplings can be as sizable as flavor conserving ones. We will include this general effect in our later discussion.

\subsection{CP Asymmetry in Bino decay}\label{secsec: CPV}
In analogy to conventional baryogenesis via late decay, a non-vanishing $\epsilon_{CP}$ requires the interference between tree-level decay of $\tilde{B}$ and a loop process, which should involve a physical CP phase in couplings as well as a kinematic cut\cite{Kolb:1990vq}. Leading 1-loop diagram requires another Majorana fermion with $\cancel{CP}$ and $\cancel{B} (\cancel{L})$ couplings in order to ensure a physical CP phase, as well as to meet the requirement from the Nanopolous-Weinberg theorem for baryogenesis\cite{Nanopoulos:1979gx}\footnote{There can be other diagrams without including another Majorana fermion such as in \cite{Cline:1990bw}, where RPV coupling and a $\cancel{CP}$ and RPV A-term appear at the loop vertices. The asymmetry from the diagrams in \cite{Cline:1990bw} thus is very sensitive to the size of RPV couplings and the size of RPV A-terms. More importantly, in the split SUSY case, the asymmetry there is highly suppressed as a result of having two heavy scalar propagators in the loop.}. Within the content of MSSM, we do have such candidates: wino or gluino. Any individual $M_i$ (i=1,2,3) can carry a complex phase, but may be rotated away. As we will elaborate later, the co-appearance of a pair of them such as having both $\tilde{B}$ and $\tilde{W} (\tilde{g})$ in the loop diagram is crucial to secure a physical CP phase, even in the case that SUSY breaking is flavor universal such that $\bf{m_{L/R,\alpha}^2}$ are proportional to real identity matrices. In split SUSY, experimental constraints on flavor violation are much looser due to the heavy scalar masses. Therefore $\bf{m_{L/R,\alpha}^2}$ can provide a sizable source of physical CP phase in addition to that from $M_i$. Meanwhile, as a result of color charge and the chirality structure of couplings, $\tilde{W}$ can only directly couple to $\cancel{L} $ operators $LL\bar{e}, LQ\bar{d}$, while $\tilde{g}$ couples to $LQ\bar{d}$ or $\bar{u}\bar{d}\bar{d}$. Furthermore, as will be shown explicitly, kinematic cuts are possible only when $M_1>M_2$ or $M_1>M_3$. In the following, we focus on analyzing two typical cases: direct baryogenesis with light gluino ($M_1>M_3$) by $\bar{u}\bar{d}\bar{d}$ coupling; leptogenesis with light wino ($M_3>M_1>M_2$) by $LQ\bar{d}$ coupling. The spectrum of the former case optimizes the amount of baryon asymmetry and viable parameter space as we will see, while the latter case is well motivated by conventional anomaly mediation models and optimizes the opportunity of bino search at hadron colliders by allowing its production from gluino cascade decays. Notice that for the leptogenesis model with $LQ\bar{d}$ coupling, we may also consider the spectrum with a light gluino ($M_1>M_3$) which would increase the amount of asymmetry due to the contribution from gluino loop which contains strong couplings.

\subsubsection{Baryogenesis with light gluino}
\begin{figure}
  \begin{minipage}{3in}
        \includegraphics[width=50mm]{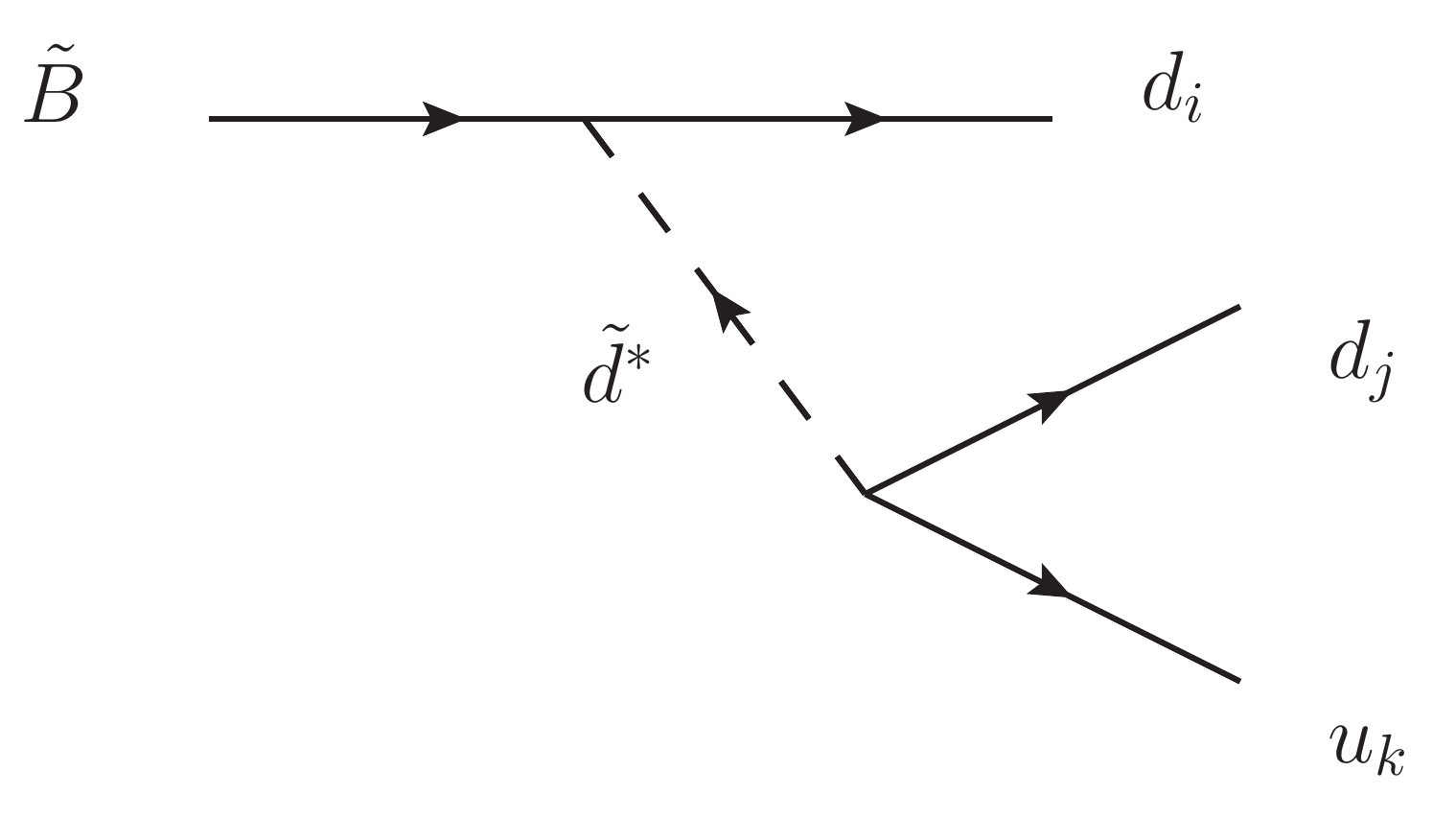} \\
        \centering{(a) }
    \end{minipage}
  \qquad
  \begin{minipage}{3in}
    \includegraphics[width=50mm]{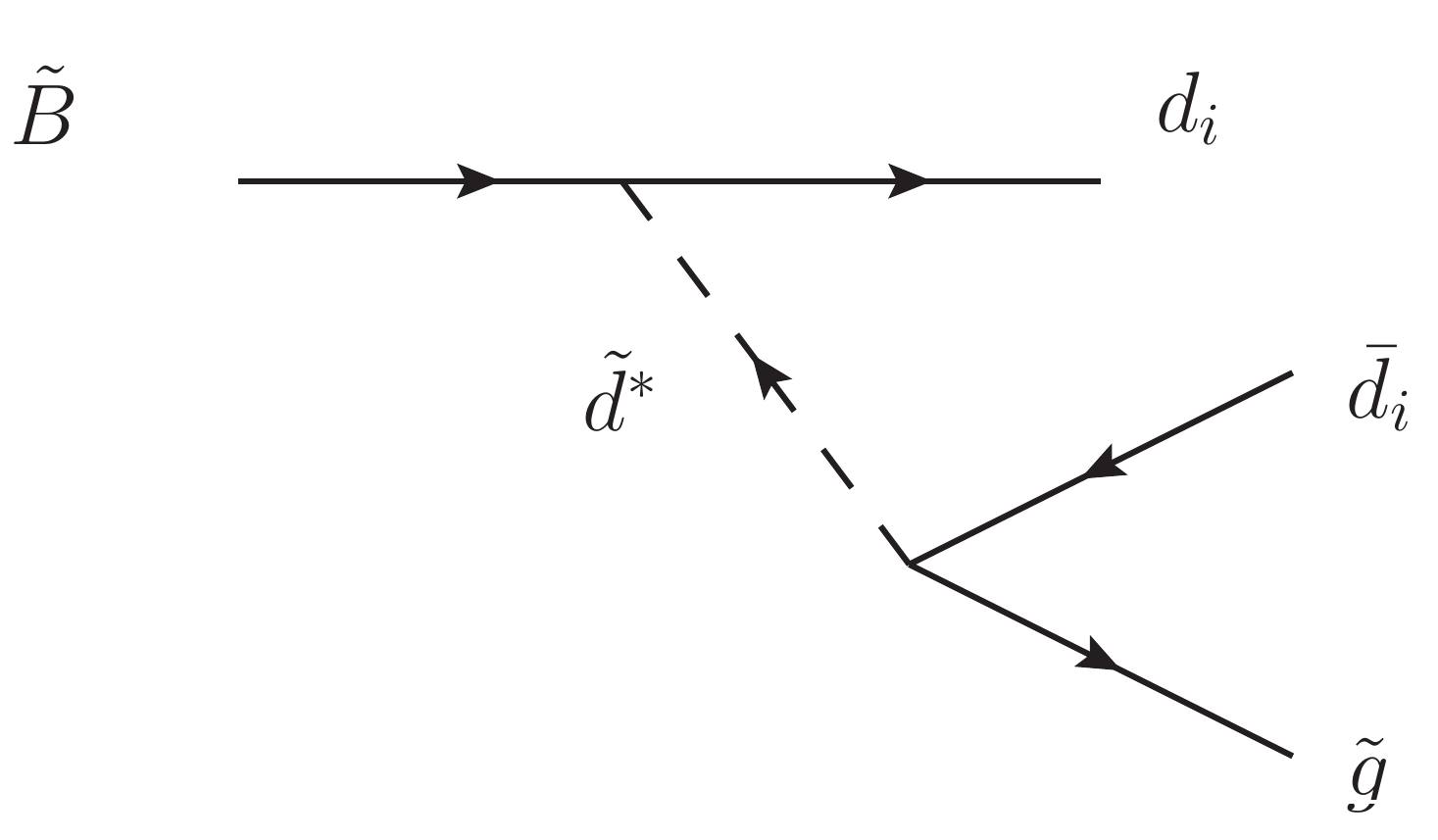}  \\
    \centering{(b)}
  \end{minipage}
  \caption{Tree-level decays of $\tb$ in the direct baryogenesis model. (a): $\cancel{B}$ decay that triggers baryogenesis; (b): $B$-conserving decay.}\label{fig:bg_treedec}
\end{figure}

\begin{figure}
  \begin{minipage}{3in}
        \includegraphics[width=60mm]{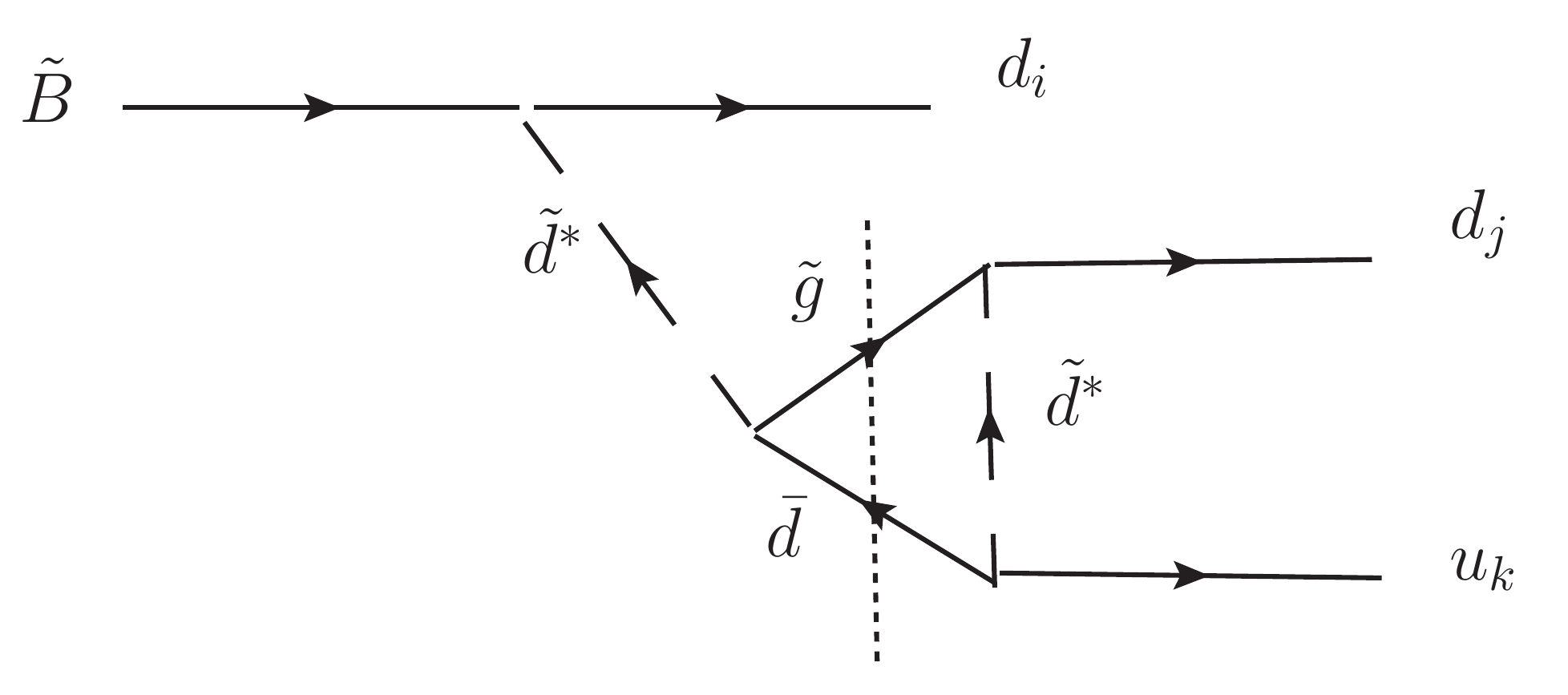} \\
        \centering{(a) }
    \end{minipage}
  \qquad
  \begin{minipage}{3in}
    \includegraphics[width=60mm]{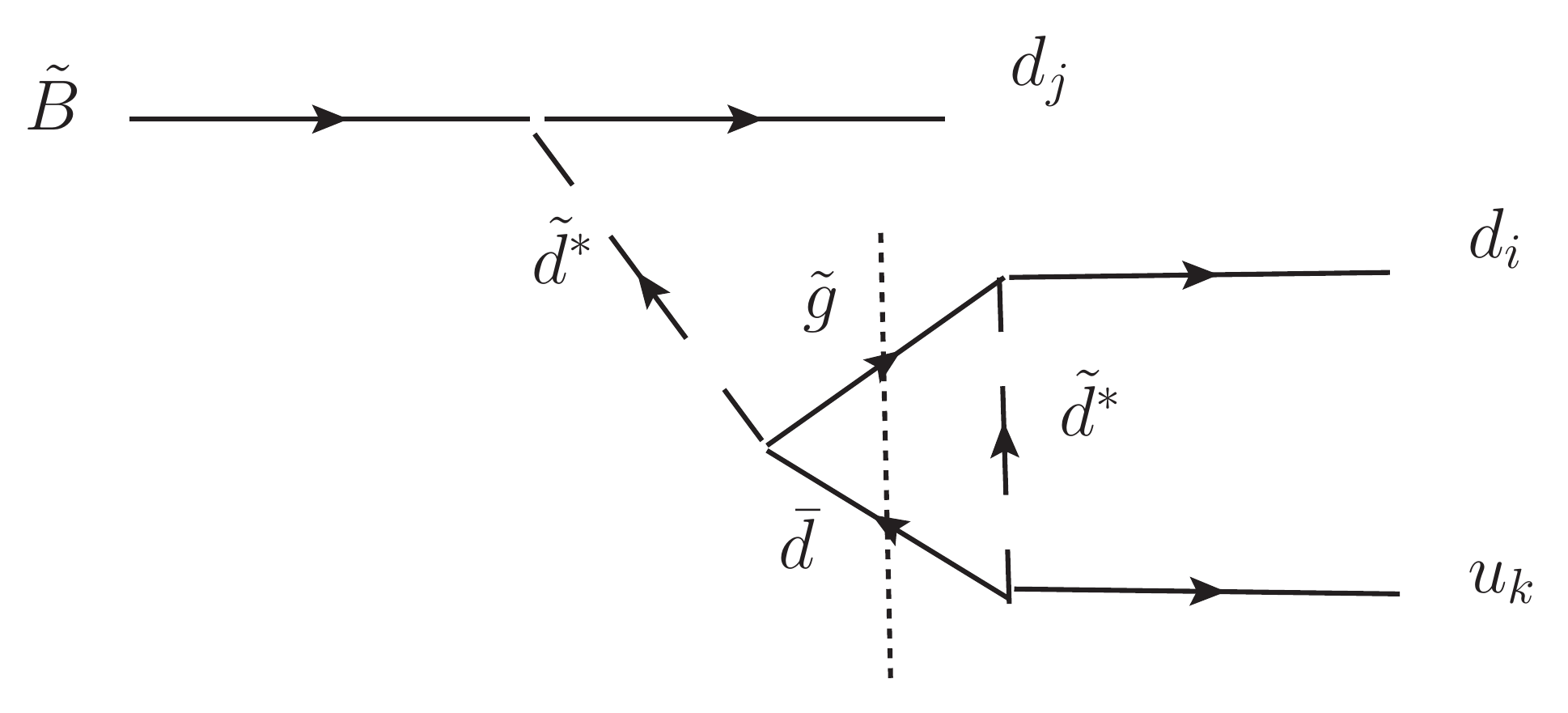}  \\
    \centering{(b)}
  \end{minipage}
  \centering{\includegraphics[width=60mm]{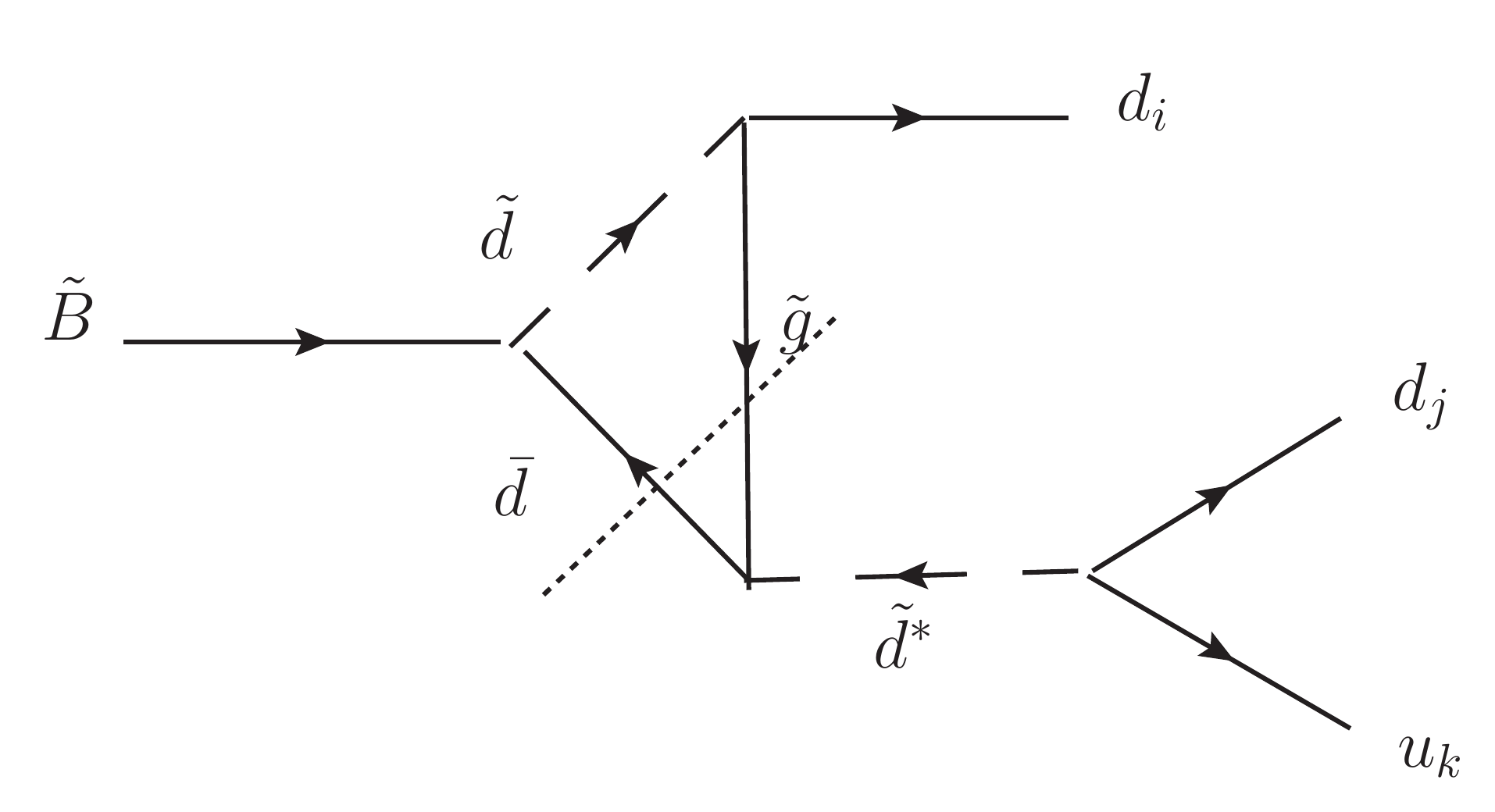}}  \\
    \centering{(c)}
  \caption{$\cancel{B}$ loop diagrams in the direct baryogenesis model. (a): does not lead to a CP asymmetry in $\cancel{B}$ $\tb$ decay. (b): contributes to CP asymmetry when the flavor and CP violation in squark mass matrices are sizable. (c): produces a CP asymmetry by interfering with Fig.\ref{fig:bg_treedec}(a) even in absence of flavor and CP violation in squark mass matrices.}\label{fig:bg_loopdec}
\end{figure}

We start with a direct baryogenesis model enabled by the $\bar{u}\bar{d}\bar{d}$ type of RPV operator. The $\cancel{B}$ tree-level decay is shown in Fig.\ref{fig:bg_treedec}(a). We then consider loop diagrams that may lead to a CP asymmetry by interference with the tree-level process. The lowest order candidate loop diagrams are shown in Fig.(\ref{fig:bg_loopdec}). Dotted lines show the position of the kinematic cut. Apparently $m_{\tilde{g}}<m_{\tilde{B}}$ is necessary to enable the cut. $\tilde{B}$ couplings can be complex by absorbing the phase $\phi$ in $M_1$. The product of the couplings from the interference between tree-level and loop graph as in Fig.\ref{fig:bg_loopdec}(a) is real after summing over generations. Under the conventional assumption of flavor universal SUSY breaking, product of couplings from interference with diagram Fig.\ref{fig:bg_loopdec}(b) is also real since the $\tilde{B}$ couplings originate from gauge couplings. However, as mentioned earlier, with split spectrum, the flavor mixing and associated phases in squark mass matrices can be large without being in conflict with experimental limits. If this is the case, diagram Fig.\ref{fig:bg_loopdec}(b) may have a non-zero imaginary part. The analogy of these two diagrams were considered in an earlier work \cite{Claudson:1983js} where they introduced two new singlet Majorana fields with general Yukawa couplings to different flavor combinations, and concluded that the SUSY embedding of their model did not work. Finally and most importantly, the diagram shown in Fig.\ref{fig:bg_loopdec}(c), which is not included in \cite{Claudson:1983js}, provides a non-vanishing CP phase even in the absence of flavor mixing and CP violation in squark mass matrices. \\
   
     Now we first examine the possible decay channels of $\tb$ before computing the CP asymmetry. Due to the uncertainties in the flavor structure of squark mass matrices, and for simplicity, in this section and onwards we will present formulae assuming flavor diagonal couplings of gauginos and focus on the contribution from Fig.\ref{fig:bg_loopdec}(c). We also make the simplest assumption that RPV couplings $\lambda^{''}_{ijk}$ takes a universal value of $\lambda^{''}$. It is straightforward to generalize them to involve other flavor patterns in the gaugino couplings and RPV couplings.
\\
 The tree level decay rate of  $\tilde{B}\rightarrow udd+\bar{u}\bar{d}\bar{d}$ via diagram in Fig.\ref{fig:bg_treedec}(a) is\footnote{As we will require $T_d<M_1$ to suppress washout, the thermal decay rate can be well approximated by the decay rate in vacuum since the dilation factor is negligible at $T_d<M_1$.}:
\beq
  \Gamma_{\tilde{B},\cancel{B}}=\frac{(\sqrt{2}\lambda''Y_dg_1)^2}{512\pi^3}\frac{m_{\tilde{B}}^5}{m_0^4}.\label{bv_decayrate}
\eeq
Notice that the above result is for one flavor combination of $udd$ operator and $\tilde{d}$ mediator. When computing the total decay width under our simplified assumption of flavor universal RPV couplings, a combinatorial factor of $18$ needs to be included. There are analogous diagrams with $\tilde{u}$ mediator, which we do not include here just for simplicity when demonstrating the working principle. The simplification can also be justified if $\tilde{u}$ is moderately heavier than $\tilde{d}$ so that the contribution from diagrams with $\tilde{u}$ are subleading. Due to the condition $m_{\tilde{g}}<m_{\tilde{B}}$, another $B$-conserving decay channel is allowed as well: $\tilde{B}\rightarrow d\bar{d}\tilde{g}$, as shown in Fig.\ref{fig:bg_treedec}(b). Here and in the later asymmetry calculation we assume $m_{\tb}$ and $m_{\tg}$ are well separated, so there is no extra kinematic suppression on the decay rate:
\bea
  \Gamma_{\tilde{B}\rightarrow d\bar{d}\tilde{g}}&=&\frac{(Y_dg_1g_3)^2}{1024\pi^3}\frac{m_{\tilde{B}}^5}{m_0^4}\label{bc_decay}
  \eea

In order to avoid additional branching ratio suppression on $\epsilon_{CP}$, we need the $B$-violating channel to be dominant over or comparable to the $B$-conserving one. Due to the larger multiplicity factor, this can be easily realized with $\lambda''\gtrsim O(0.1)$.\\

Now we move on to compute the $\epsilon_{CP}$ from the interference with diagram Fig.\ref{fig:bg_loopdec}(c):
\bea
\Delta{\Gamma_{\tilde{B},\cancel{B}}}&\equiv&\Gamma_{\tilde{B}\rightarrow udd}-\Gamma_{\tilde{B}\rightarrow \bar{u}\bar{d}\bar{d}}=\frac{Im \left[(\sqrt{2}Y_dg_1g_3\lambda'')^2e^{i\phi}\right]C_2}{10240\pi^4}\frac{m_{\tb}^7}{m_0^6}\\
\epsilon_{CP}&\equiv&\frac{\Delta{\Gamma_{{\tilde{B},\cancel{B}}}}}{\Gamma_{{\tilde{B}}}}=\frac{g_3^2 Im[e^{i\phi}]C_2}{20\pi}\frac{m_{\tb}^2}{m_0^2},\label{epsiloncp_bg}
\eea
where $C_2=\frac{4}{3}$ is the quadratic Casimir from $\tg$ vertices. The 2nd line in eq.(\ref{epsiloncp_bg}) assumes $\Gamma_{\tilde{B}}\approx\Gamma_{\tilde{B},\cancel{B}}$. When exploring parameter space in later numerical study we include the contribution to the total width from other decay channel (eq.(\ref{bc_decay})). As can be seen from eq.(\ref{epsiloncp_bg}), there is an additional mass suppression factor in addition to 1-loop factor. To give a quick numerical sense: with $O(1)$ CP phase in $M_1$, and $m_{\tb}\sim$ TeV, $m_0\sim100$ TeV, we find $\epsilon_{CP}\sim10^{-6}$. The asymmetry given in eq.(\ref{epsiloncp_bg}) only includes the contribution from Fig.\ref{fig:bg_loopdec}(c) assuming flavor diagonal bino couplings. With possible flavor and CP violations in squark mass matrices included, it is reasonable to expect the asymmetry gets enhanced by a factor of 2 or more numerically.

\subsubsection{Leptogenesis with light wino}
\begin{figure}
  \centering{\includegraphics[width=50mm]{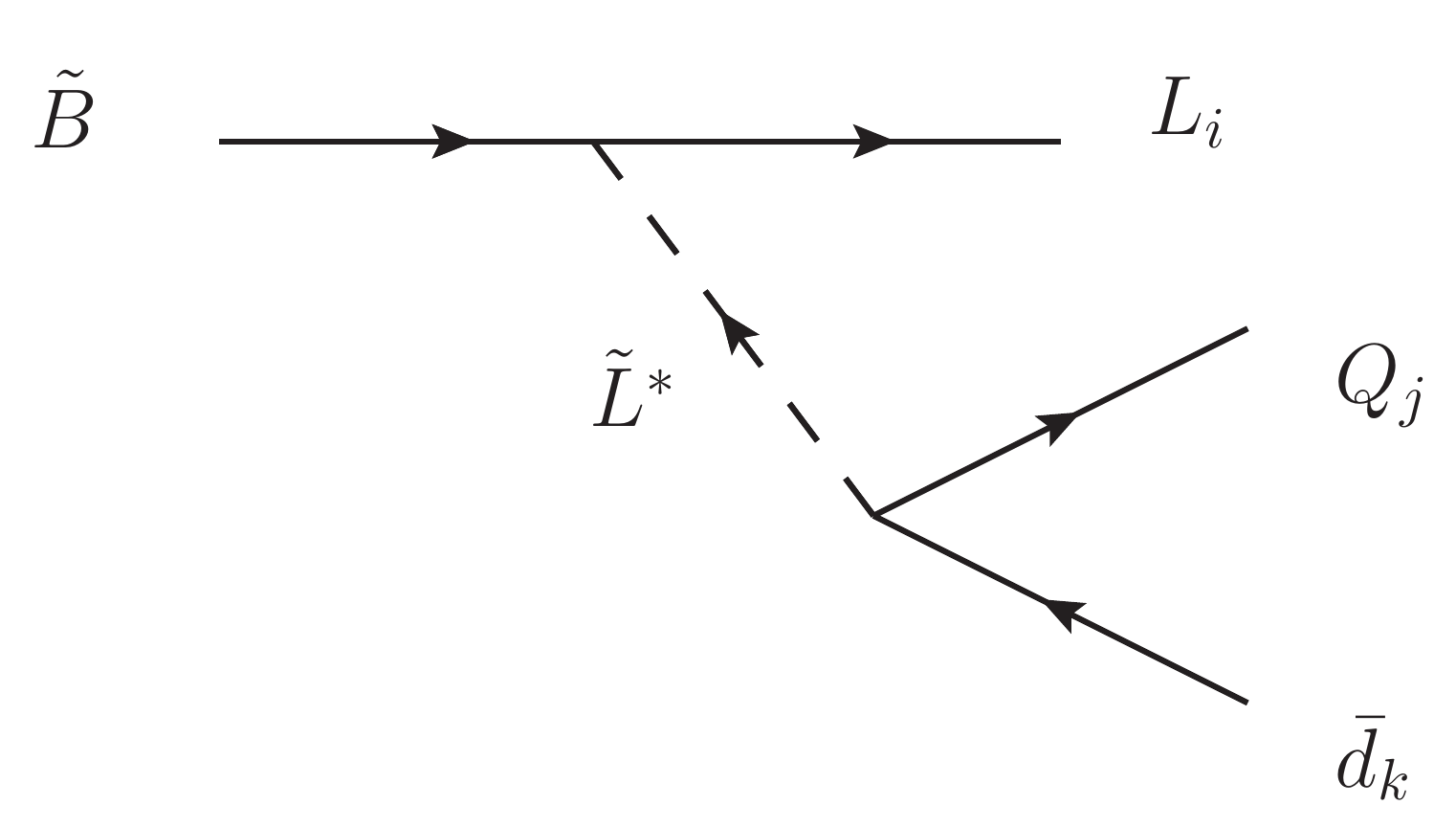}}  \\
    \centering{(a)}\\
  \begin{minipage}{3in}
        \includegraphics[width=50mm]{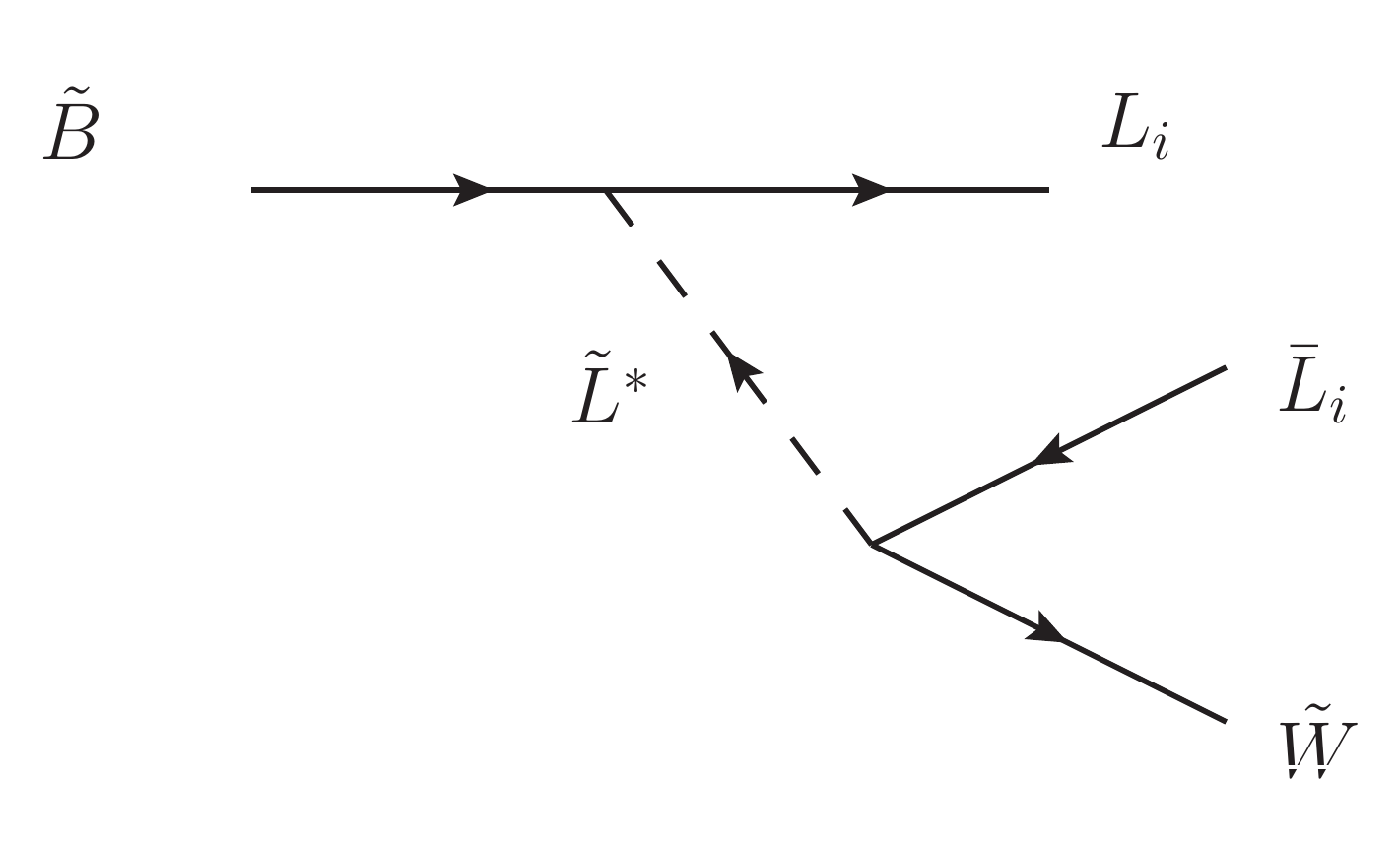} \\
        \centering{(b) }
    \end{minipage}
  \qquad
  \begin{minipage}{3in}
    \includegraphics[width=60mm]{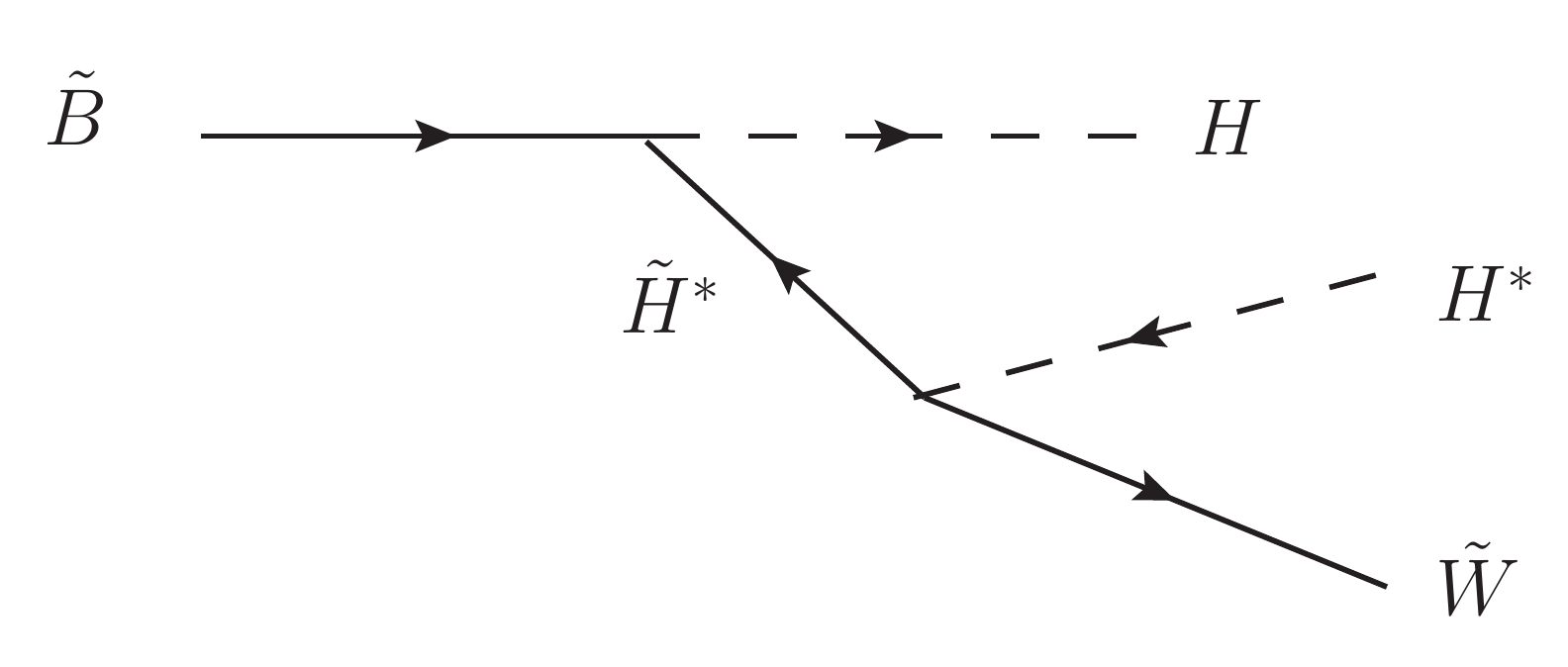}  \\
    \centering{(c)}
  \end{minipage}
  \caption{Tree-level decays of $\tb$ in leptogenesis model. (a): $\cancel{L}$ decay that triggers leptogenesis; (b), (c): $L$-conserving decay.}\label{fig:lg_treedec}
\end{figure}

\begin{figure}
  \begin{center}
   \begin{minipage}{3in}
   \includegraphics[width=60mm]{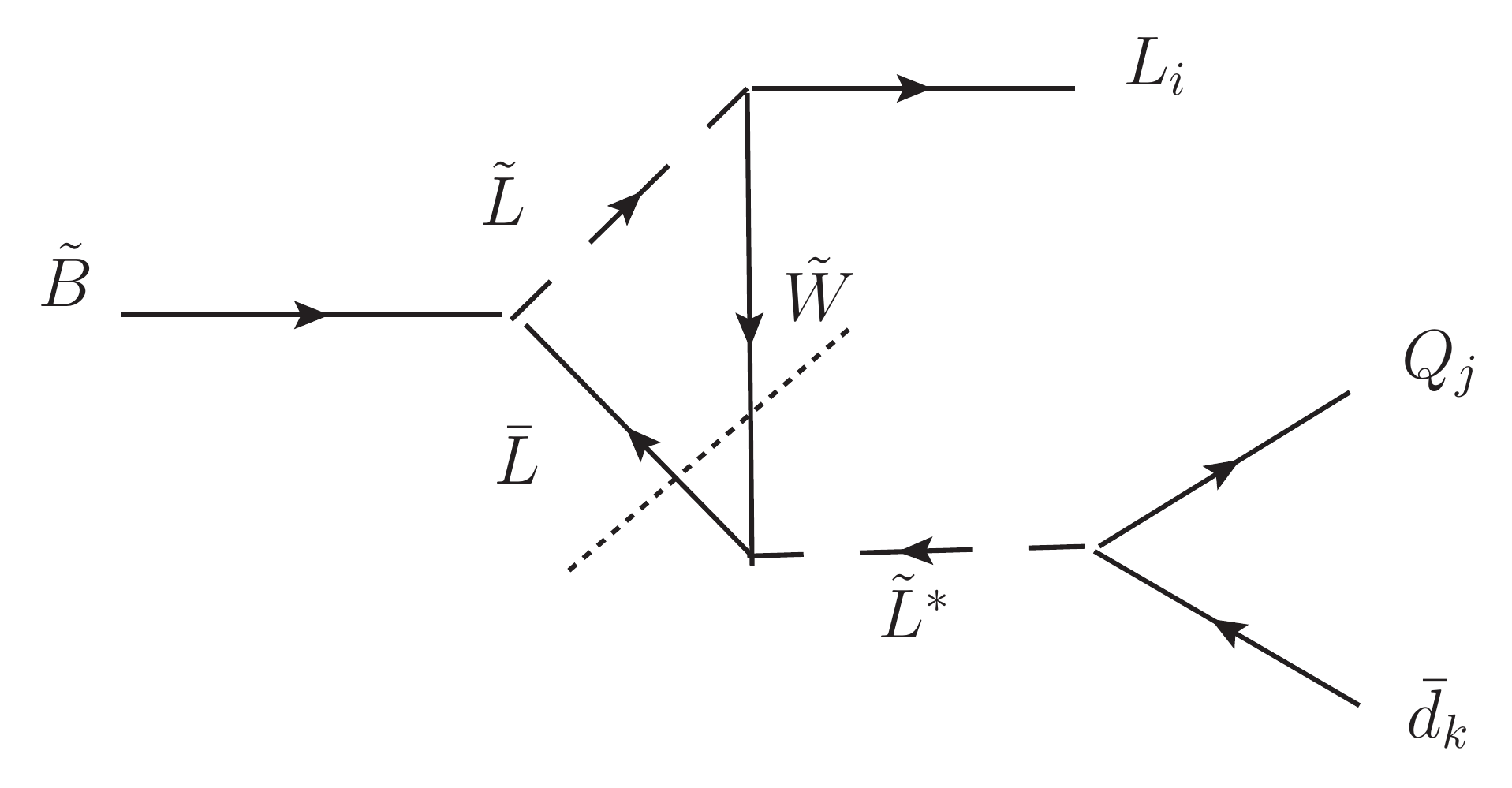} \\
   \centering{(a)}
   \end{minipage}
   \begin{minipage}{3in}
   \includegraphics[width=65mm]{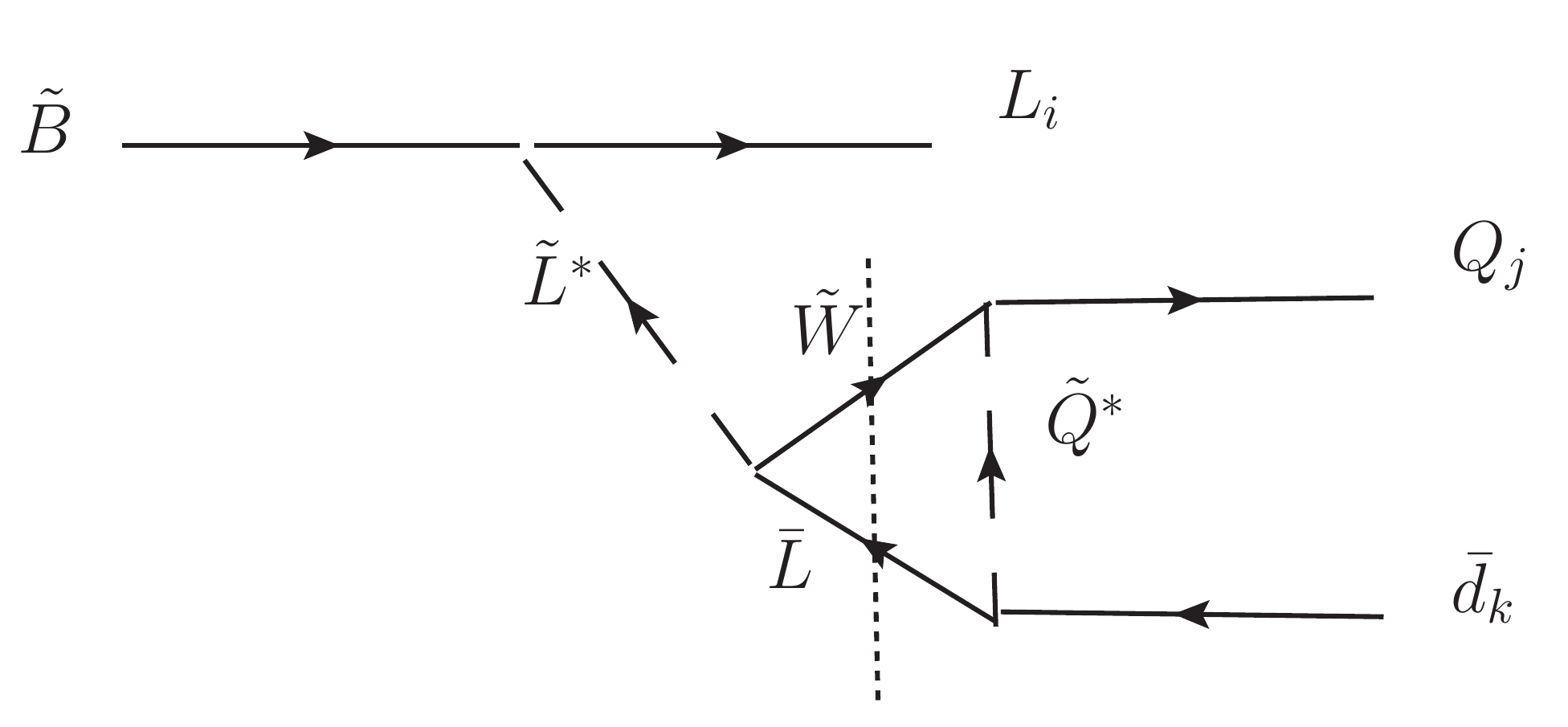} \\
   \centering{(b)}
   \end{minipage}
  \caption{$\cancel{L}$ loop diagrams in leptogenesis model. (a) produces a CP asymmetry by interference with Fig.\ref{fig:lg_treedec}(a) even in absence of flavor and CP violation in sfermion mass matrices. (b) contributes to CP asymmetry when the flavor and CP violation in sfermion mass matrices are sizable.}\label{fig:lg_loopdec}
  \end{center}
\end{figure}

In analogy to the direct baryogenesis model, for the leptogenesis model we focus on tree-level $\tb$ decay as shown in Fig.\ref{fig:lg_treedec}(a) and its interference with the loop diagram shown in Fig.\ref{fig:lg_loopdec}(a) which gives rise to a CP asymmetry. The interference with Fig.\ref{fig:lg_loopdec}(b) can give comparable additional asymmetry with sizable flavor and CP violation in slepton mass matrices. The tree level decay rate of  $\tilde{B}\rightarrow LQ\bar{d}+\bar{L}\bar{Q}d$ via the diagram in Fig.\ref{fig:lg_treedec}(a) is:
\beq
  \Gamma_{\tilde{B},\cancel{L}}=\frac{(\sqrt{2}\lambda'Y_Lg_1)^2}{512\pi^3}\frac{m_{\tilde{B}}^5}{m_0^4}.\label{lv_decayrate}
\eeq
Again the above result is for one flavor combination of $LQ\bar{d}$, and focuses on $\tilde{L}$ mediator for simplicity. When computing the total decay width, under our simplified assumption of flavor universal RPV couplings, a factor of 27 needs to be included to account for the flavor multiplicity. Due to the condition $m_{\tilde{W}}<m_{\tilde{B}}$, two other $L$-conserving decay channels open up: $\tilde{B}\rightarrow L\bar{L}\tilde{W}$, $\tilde{B}\rightarrow H^*H\tilde{W}$, as shown in Fig.\ref{fig:lg_treedec}(b),(c). Assuming $m_{\tb}$ and $m_{\tw}$ are well separated, the decay rates are as follows:
\bea
  \Gamma_{\tilde{B}\rightarrow L\bar{L}\tilde{W}}&=&\frac{(Y_Lg_1g_2)^2}{3072\pi^3}\frac{m_{\tilde{B}}^5}{m_0^4}\label{lc_decay1}\\
  \Gamma_{\tilde{B}\rightarrow H^*H\tilde{W}}&=&\frac{(Y_{H}g_1g_2)^2}{384\pi^3}\frac{m_{\tilde{B}}^3}{\mu^2}\label{lc_decay2}
\eea
Analogous to the baryogenesis case, the channel $\tilde{B}\rightarrow L\bar{L}\tilde{W}$ is subleading compared to the $\tilde{B}$ channel provided that $\lambda'\gtrsim O(0.1)$. The rate of the channel $\tilde{B}\rightarrow H^*H\tilde{W}$ can be subdominant if $\mu\gg m_0$. This is consistent with the already existing requirement of $\mu\gg m_0$ for getting enough $\Omega_{\Delta B}$ as discussed earlier. 
With $\cancel{L}$ decay dominating, the $\epsilon_{CP}$ from the interference with diagram Fig.\ref{fig:lg_loopdec}(a) is:
\bea
\Delta{\Gamma_{\tilde{B},\cancel{L}}}&\equiv&\Gamma_{\tilde{B}\rightarrow LQ\bar{d}}-\Gamma_{\tilde{B}\rightarrow \bar{L}\bar{Q}d}=\frac{Im \left[(\sqrt{2}Y_Lg_1g_2\lambda')^2e^{i\phi}\right]}{10240\pi^4}\frac{m_{\tb}^7}{m_0^6}\\
\epsilon_{CP}&\equiv&\frac{\Delta{\Gamma_{\tilde{B},\cancel{L}}}}{\Gamma_{\tilde{B}}}=\frac{g_2^2 Im[e^{i\phi}]}{20\pi}\frac{m_{\tb}^2}{m_0^2}.\label{epsiloncp_lg}
\eea
The 2nd line in eq.(\ref{epsiloncp_lg}) assumes $\Gamma_{\tilde{B}}\approx\Gamma_{\tilde{B},\cancel{L}}$. We include the contribution to the total width from other decay channels (eq.(\ref{lc_decay1},\ref{lc_decay2})) when exploring parameter space in our numerical study.

\section{Computation of $\Omega_{\Delta B}$, Constraints}\label{sec:results}
\begin{figure}
  \centering{\includegraphics[width=50mm]{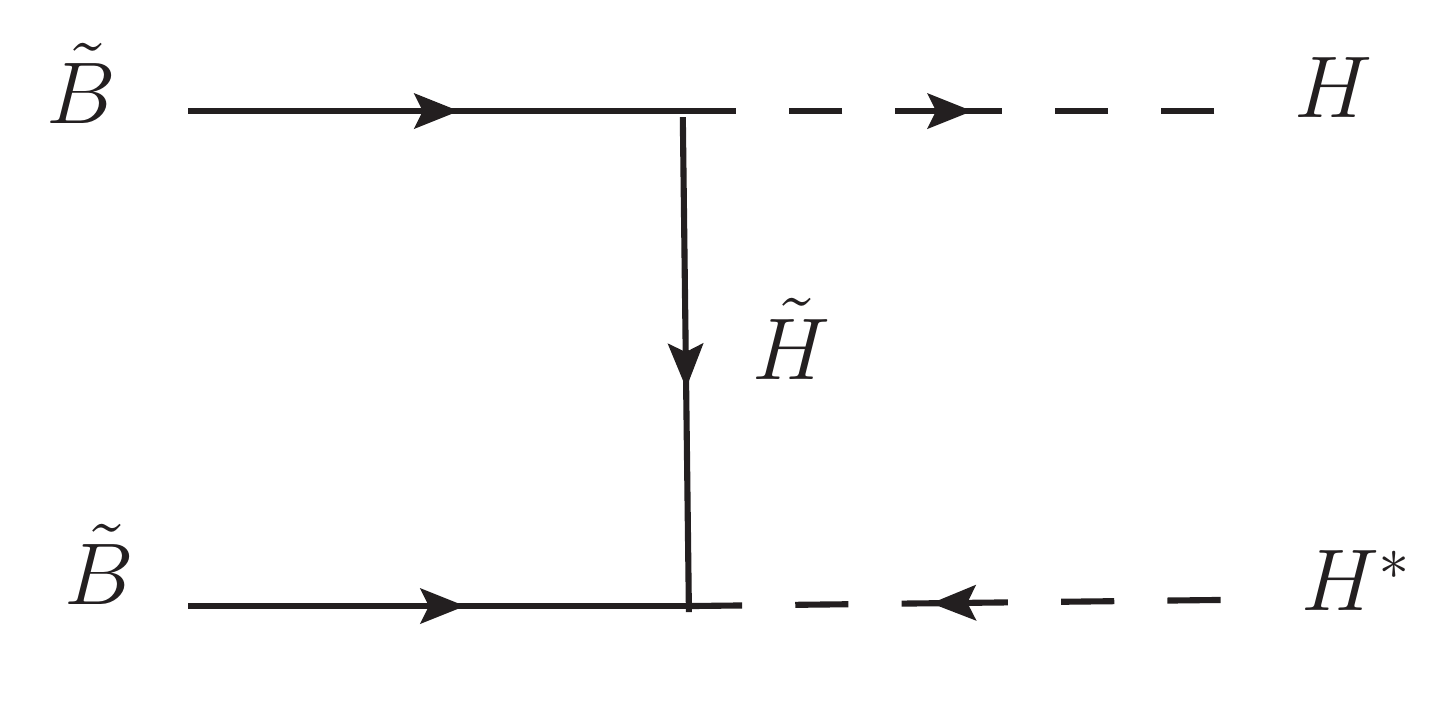}}  \\
    \centering{(a)}\\
  \begin{minipage}{3in}
        \includegraphics[width=50mm]{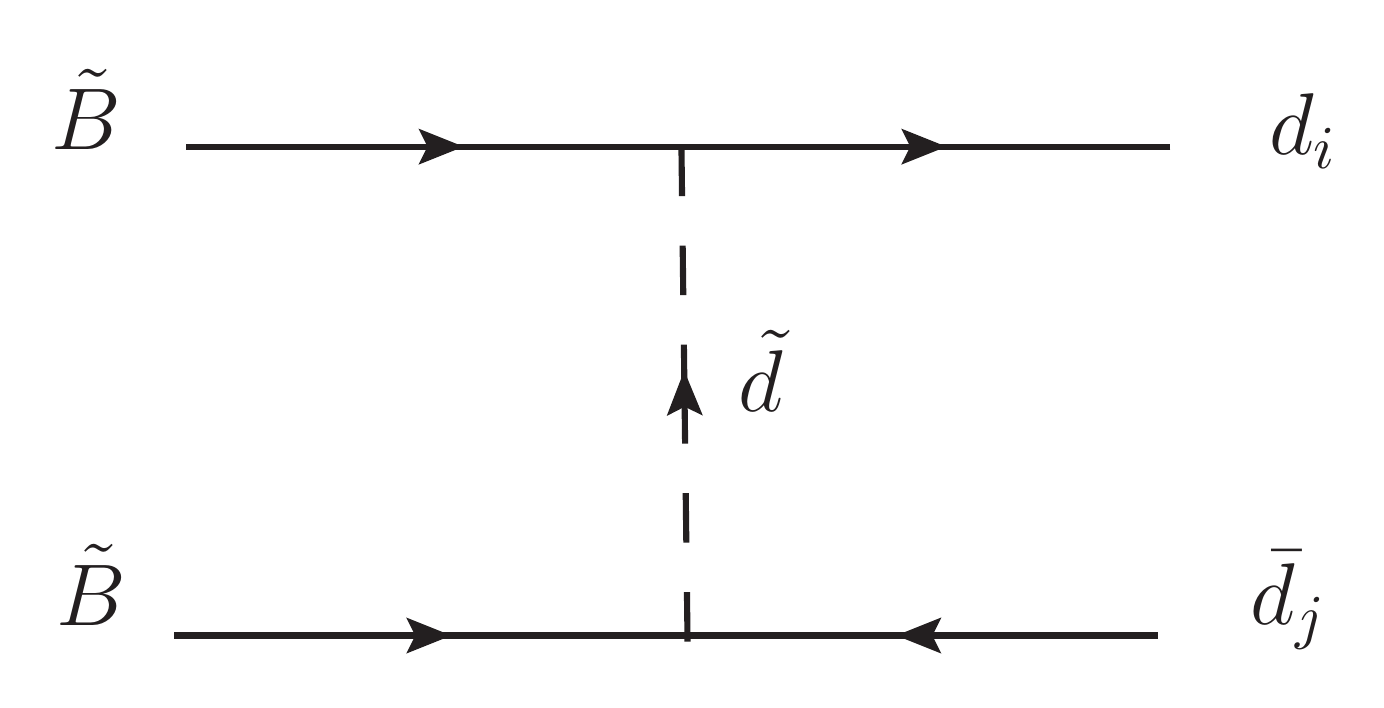} \\
        \centering{(b) }
    \end{minipage}
  \qquad
  \begin{minipage}{3in}
    \includegraphics[width=50mm]{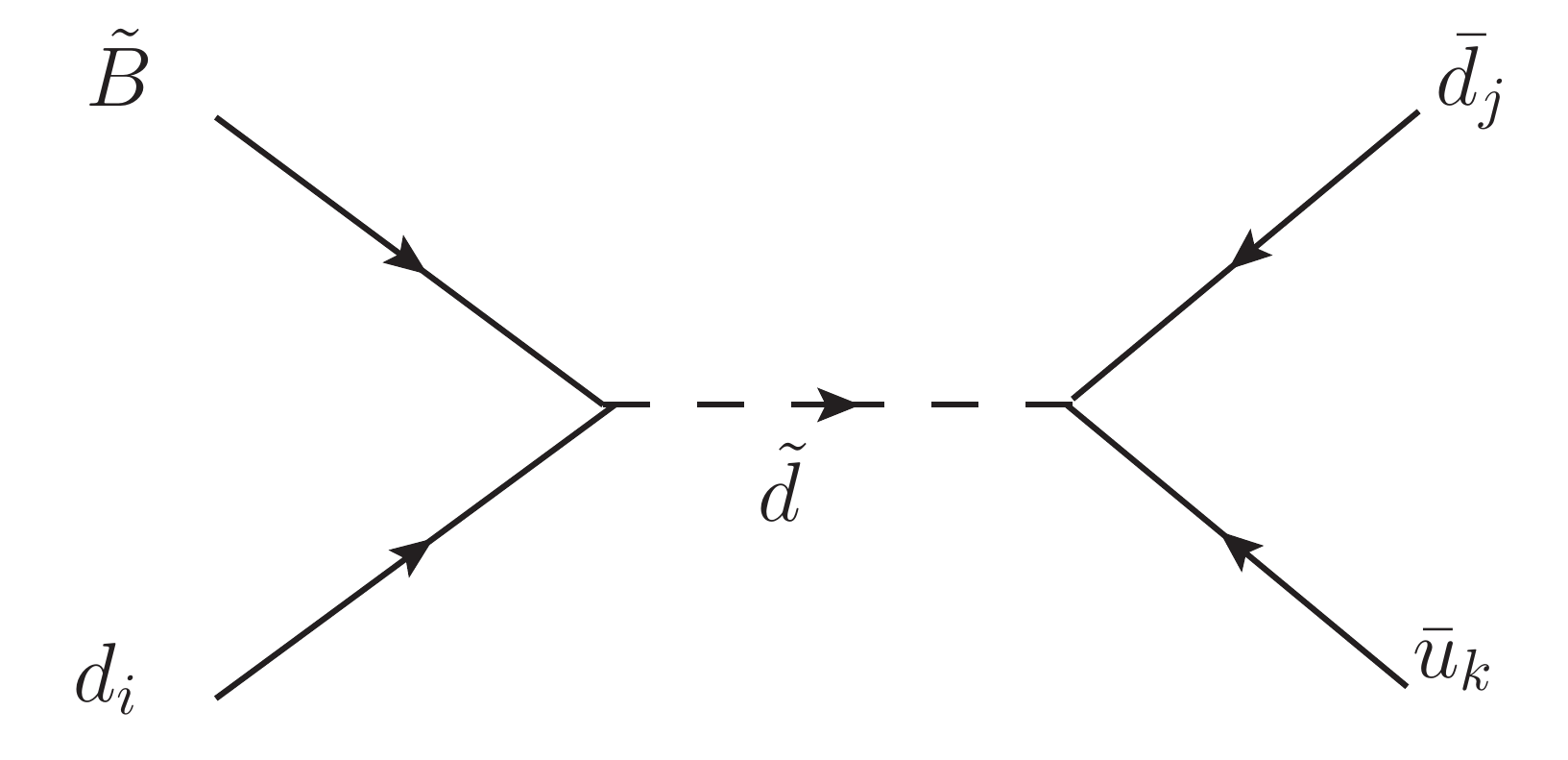}  \\
    \centering{(c)}
  \end{minipage}
  \caption{(a): Leading annihilation process of $\tb$; (b), (c): examples of other annihilation channels.}\label{fig:ann}
\end{figure}

  \indent
  In this section we compute $\Omega_{\Delta B}$ in detail, and present numerical results taking into account of all of the cosmological constraints. We start with analyzing the thermal annihilation of a TeV $\tb$ and its would-be relic abundance $\Omega^{\tau\rightarrow\infty}_{\tb}$. In the case of the leptogenesis the decay needs to occur after thermal freezeout time $T_f$ while before electroweak phase transition (EWPT) around $T_c\approx100$ GeV, so that the sphaleron process is still efficient to transfer the asymmetry to baryons. In the case of baryogenesis, in principle the decay can happen well after EWPT. However, as we will show in the numerical results, due to the suppressed $\epsilon_{CP}$ the requirement of obtaining sufficient $\Omega_{\Delta B}$ typically push $T_f$ higher, before EWPT. Even when $\tb$ decay after EWPT, the process we discuss here remains the leading contribution, and the change in results is expected to be $O(1)$ or less. Before EWPT $\tb$ is a pure bino without mixing with wino or higgsino. Higgsinos at this stage are pure Dirac: $\tilde{H}=(\tilde{H}_u,\tilde{H}_d)$. The requirement of $\mu\gg m_0$ implies $\mu^2\approx B_\mu$ in order to have realistic electroweak symmetry breaking. Consequently, the rotation angle $\alpha$ rotating from gauge basis $(H_u, H_d^*)$ to mass basis $(H,H')$ is $\alpha\approx \pi/4$, where the very light mode $H$ relates to the SM higgs boson after EW symmetry breaking. The hierarchical spectrum 100 GeV$\sim m_H\ll m_{H'}$ results from the split mass scales and fine cancelation. The major diagram that contributes to $\tb$ annihilation before EWPT is $\tb\tb\rightarrow HH^*$ as shown in Fig.\ref{fig:ann}(a). At this stage the light Higgs spectrum consists of a complex doublet $H$ with four real degrees of freedom. The annihilation cross section is given by:
  \beq
    \sigma_{HH^*}(s)=\frac{g_1^4}{32\pi} \frac{s-4M_1^2}{s\sqrt{1-4M_1^2/s}}\frac{1}{\mu^2}.
  \eeq
The thermally averaged cross-section is:
\bea
  \langle\sigma_{HH^*}v\rangle=\frac{1}{8M_1^4TK_2^2(M_1/T)}\int^\infty_{4M_1^2}ds \sigma_{HH^*}(s)(s-4M_1^2)\sqrt{s}K_1(\frac{\sqrt{s}}{T})
  \eea

In a large range of $(\mu, m_s)$ parameter space including part of the $\mu\gg m_s$ region of our interest, the above process ${\tb}{\tb}\rightarrow HH^*$ is the dominant annihilation channel. However, with $m_s$ fixed at a finite value while keep increasing $\mu$, at some point other annihilation processes such as those shown in Fig.\ref{fig:ann}(b,c) can become comparable or dominate over the channel of ${\tb}{\tb}\rightarrow HH^*$. In particular, in the limit $\mu\rightarrow\infty$ the process ${\tb}{\tb}\rightarrow HH^*$ would decouple. Therefore, to cover the most general parameter space, we should include the other annihilation channels in our analysis as well. Among these other annihilation channels we expect the $\cancel{B}(\cancel{L})$ channel of type $\tb d\rightarrow \bar{u}\bar{d}$ ($\tb L\rightarrow \bar{Q}d$) to be dominant due to its large multiplicity factor and the lack of p-wave suppression. The analogy of these additional annihilation channels are discussed in \cite{Claudson:1983js, Cheung:2013hza} where the baryogenesis models are based on two general singlet Majorana fermions and a new heavy diquark scalar. Under our simplified assumption of universal RPV couplings, the thermal cross section of $\cancel{B}$ annihilation (Fig.\ref{fig:ann}(c)) is:
\beq
  \langle \sigma_{\cancel{B} (\cancel{L})}v\rangle\simeq\frac{\xi^2}{10\pi}\frac{M_1^2}{m_s^4}\left[5\frac{K_4(M_1/T)}{K_2(M_1/T)}+1\right],\label{ann2}
\eeq
where $\xi=\sqrt{2}g_1Y_d\lambda^{''}, \sqrt{2}g_1Y_L\lambda^{'}$ in the case of baryogenesis and leptogenesis respectively.
 We define the freezeout temperature by solving
\beq
\Gamma_{\rm ann}(x_f)\equiv n_{\rm eq}(x_f)\langle\sigma v_{\rm ann}(x_f)\rangle=H(x_f),
\eeq
where $x_f\equiv M_1/T_f$, $\langle\sigma v_{\rm ann}\rangle$ includes the sum of the $\tilde{H}$ mediated and the $\cancel{B}(\cancel{L})$ annihilations. \\
To a good approximation to the solution obtained by solving Boltzmann equation, the ``would-be'' relic abundance of bino is given by:
\bea
\nonumber
  \Omega_{\tb}^{\tau\rightarrow\infty}&=&\frac{Y_{\rm eq}(x_f)M_1}{(\rho_c/s)_0}\\
  &\simeq&\frac{2\cdot10^9~{\rm GeV}^{-1}~x_f}{g_*^{\frac{1}{2}}M_{\rm pl}\langle\sigma v_{\rm ann}(x_f)\rangle},\label{binorelic}
\eea
where $Y^{\rm eq}_{\tb}(x_f)\equiv \frac{n_{\rm eq}(x_f)}{s_(x_f)}$ is the co-moving number density of bino at $T_f$. $(\rho_c/s)_0\approx3.6h^2\cdot10^{-9}\rm GeV$ ($h\approx0.7$), where $(\rho_c)_0$ is the critical co-moving density today, $s_0$ is the current day entropy. $g_*$ counts the effective relativistic degrees of freedom, $g_*\approx100$ before EWPT. As we will see, due to the suppressed asymmetry, $x_f\lesssim5$ in most of the working parameter space.\\

Finally we obtain baryon relic abundance by combining eqs.(\ref{omegaB},\ref{epsiloncp_bg},\ref{epsiloncp_lg},\ref{binorelic}). With $\tilde{H}$ mediated annihilation dominating, we can parametrize an estimate as:
\beq
  \Omega_{\Delta B}\sim10^{-2}\left(\frac{m_{\tb}}{1\rm~ TeV}\right)\left(\frac{\mu}{10m_0}\right)^2.\label{omegab_est}
\eeq
 The above estimate is based on the baryogenesis model. With the same mass parameters, the numerical value for the leptogenesis model is expected to be up to 1 order of magnitude smaller, due to smaller gauge coupling $g_2$ (vs.~$g_3$) and smaller sphaleron distribution factor, although as mentioned earlier the asymmetry can be larger in the leptogenesis model if we consider a spectrum with $M_1>M_3$ instead. Apparently for a TeV mass bino, a split between $\mu$ and $m_0$ is necessary to obtain the observed $\Omega_{\Delta B}$. We shall comment that for fixed $m_{\tb}$ and $m_0$, $\Omega_{\Delta B}$ does not keep increasing as we increase $\mu$. For one, as discussed earlier, at some point with $\mu\gg m_0$ other annihilation channels mediated by sfermions would dominate the total cross section and limit the growth of $\Omega_{\Delta B}$. Meanwhile, for $m_{\tb}\sim O(\rm TeV)$, $\tb$ freezes out as a hot relic ($x_f\lesssim1$) at $\mu\sim10^9$ GeV and $m_0\sim10^7$ GeV, where $Y^{eq}_{\tb}(x_f)$ would saturate to its maximum value $\sim10^{-3}$, without further growth when $\langle\sigma v_{\rm ann}\rangle$ further reduces and $x_f$ becomes smaller. \\
    
    There are several potential suppression factors on top of the above estimate where we assume 100\% baryogenesis efficiency. Firstly, the asymmetry may be suppressed if $\cancel{B}(\cancel{L})$ decay has a small branching ratio (BR), compared to the $B$(L)-conserving decay channel. As discussed in Section.\ref{sec:model}, with a sizable RPV coupling ($\gtrsim O(0.1)$) and the existing condition of $\mu\gg m_0$, $\cancel{B}(\cancel{L})$ decay is typically a leading channel, and the BR suppression is of no concern. Secondly, there are $\cancel{B}(\cancel{L})$ washout processes from inverse decay or scatterings such as $\tb+u\rightarrow \bar{d}\bar{d}$, which may reduce the efficiency.  Ref.\cite{Cui:2012jh} includes a detailed discussion about these washout effects that can apply here as well. The rule of thumb is that for decay temperature $T_d<M_1$ these washout processes are suppressed by Boltzmann factors. Finally, the late decay of a massive particle may raise the concern about possible dilution of $\Omega_{\Delta B}$ from reheating effect. Assuming $\tb$ decays at temperature $T_d$ with co-moving number density $Y_{\tb}$, and the temperature of the universe after its decay is $T'$, by energy conservation, we find that $T'^4\approx T_d^4+\frac{4}{3}Y_{\tb}M_1T_d^3$. Therefore, if $Y_{\tb}M_1\ll T_d$, we have $T'\approx T_d$, and so the dilution is negligible; if $Y_{\tb}M_1\gg T_d$, $\Omega_{\Delta B}$ gets a dilution factor $\sim\frac{3T_d}{4M_1Y_{\tb}}$. The maximal value of $Y_{\tb}$ $\sim10^{-3}$ when $\tb$ freezes out with $x_f\ll1$. Thus the dilution may be significant only for very late decay when $T_d\ll M_1$, and with large $Y_{\tb}$. Meanwhile, later decay requires larger $m_0$ and smaller RPV couplings, which suppress the amount of asymmetry produced. We include such dilution factor in our numerical results, and find that it is in fact safely negligible in the parameter space giving sufficient $\Omega_{\Delta B}$.\\
   
\begin{figure}
  \begin{minipage}{3in}
        \includegraphics[width=80mm]{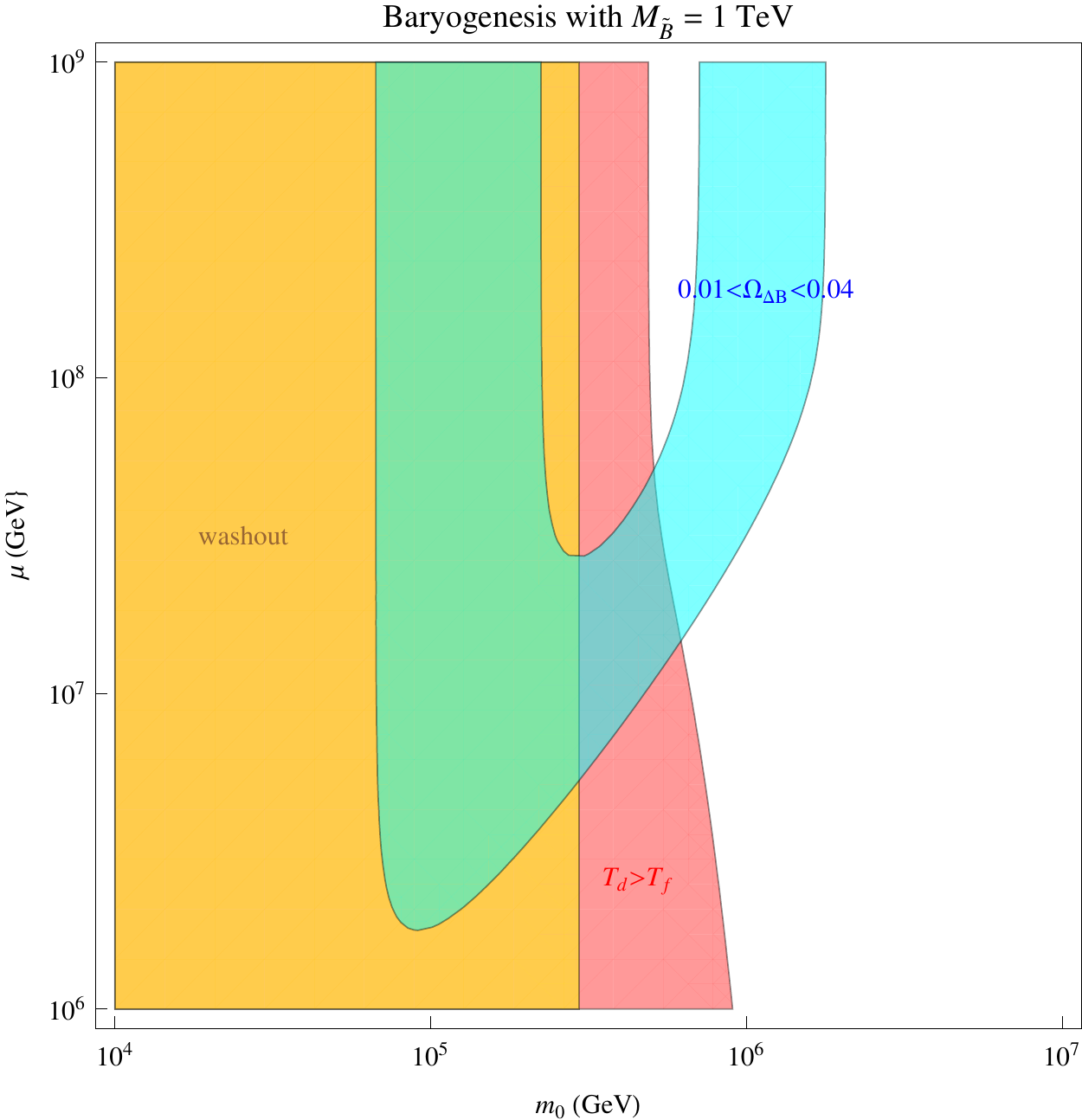} \\
        \centering{(a)}
   \end{minipage}
  \qquad
 \begin{minipage}{3in}
    \includegraphics[width=80mm]{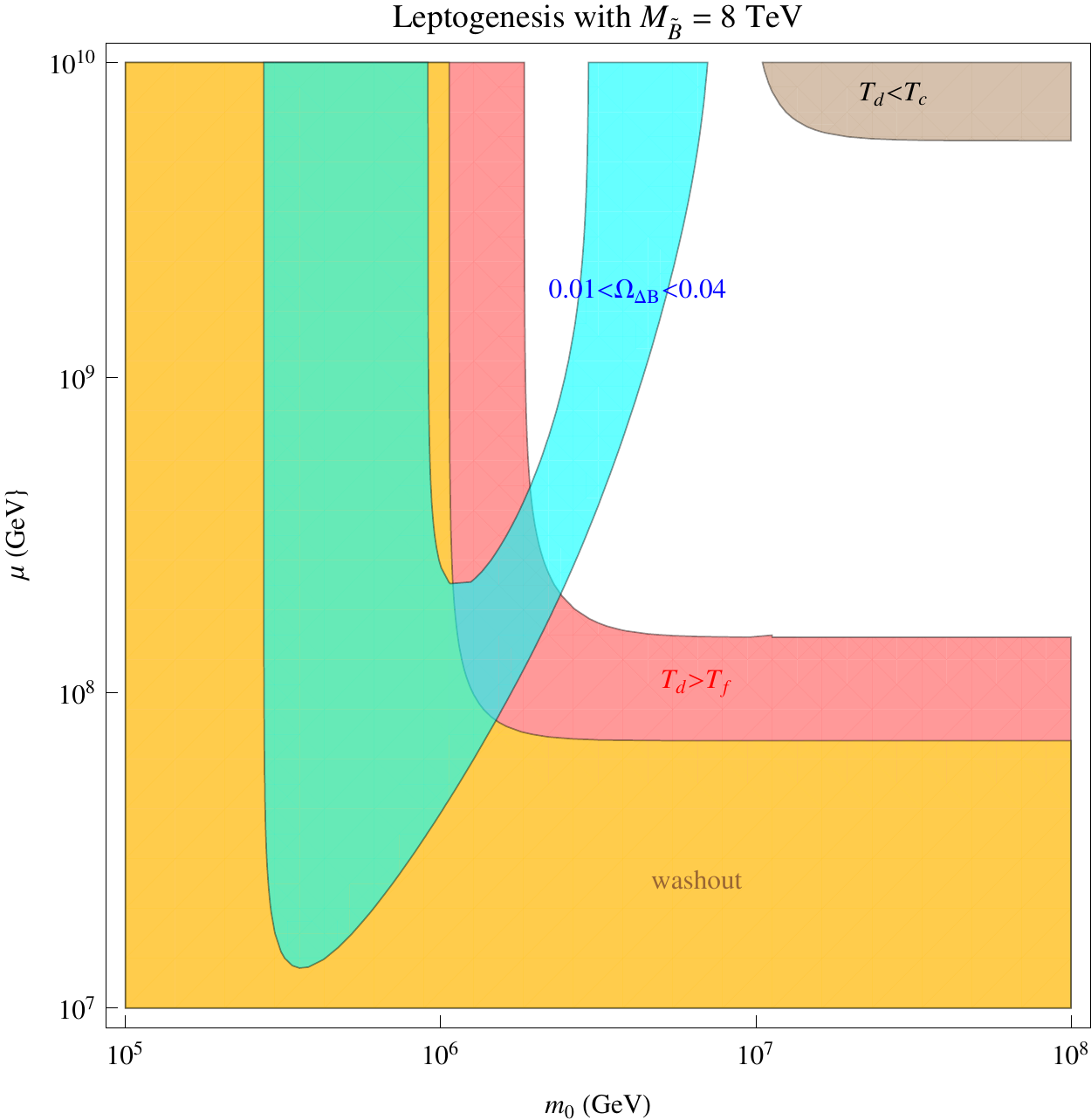}  \\
    \centering{(b)}
 \end{minipage}
 \caption{Cosmologically allowed regions of parameter space for (a) baryogenesis and (b) leptogenesis models. We set RPV couplings $\lambda^{''}=\lambda^{'}=0.2$, $\phi=\frac{\pi}{2}$. Cyan region provides baryon abundance $10^{-2}<\Omega_{\Delta B}<4\cdot10^{-2}$. In the case of leptogenesis the brown region is excluded by decay after EWPT at $T_c\approx100\rm ~GeV$. The pink region is excluded by our simple basic assumption that bino decays after freezeout. Yellow region is excluded by requiring that washout processes are suppressed ($T_d<M_{\tb}$). Yellow region is in fact all included in the pink region (so appear to be orange in the overlapped region). } \label{fig:results}
  
\end{figure}

We present the numerical results in Fig.(\ref{fig:results}), including the above possible suppression factors and other cosmological constraints we have discussed. We want to comment that the numerical analysis here focuses on presenting examples for the working principle, rather than a comprehensive precise parameter scan. The cyan region outside the yellow/pink/brown shades is viable.
The viability of the pink band with $T_f<T_d<M_{\tb}$ is in fact not strictly excluded: in this region the baryogenesis occurs before freezeout, thus the neat robust solution for $\Omega_{\Delta B}$ based on our simple assumption ($T_d<T_f$) as given in eq.(\ref{omegaB}) does not apply; solving coupled Boltzmann equations involving annihilation, decay and washout is necessary, which is analogous to the situation in the ``WIMPy baryogenesis'' scenario proposed in \cite{Cui:2011ab}. As shown in Fig.(\ref{fig:results}), compared to the leptogenesis, the baryogenesis model can achieve a similar amount of asymmetry with lighter $\tb$ and smaller splitting between $\mu$ and $m_0$, which is expected from our earlier analytic estimate. Nonetheless, as discussed earlier, asymmetry from leptogenesis may be enhanced if we consider the alternative spectrum with $M_1>M_3$, so that gluino loop may bring in larger contribution. As can be seen, for a TeV bino, in the viable parameter space, $m_0$ typically takes values of $10^2-10^4$ TeV, quite interestingly being the ``mini-split'' regime, yet independently motivated by cosmological conditions. There can be other potential constraints from particle physics such as Higgs mass and flavor physics experiments.  Higgs mass does not provide a direct constraint, since its precise value depends on or can be accommodated by other factors not directly related to our cosmological focus here, such as $A$-terms or introducing new particles. Nonetheless, as is well known, with the mini-split spectrum, it is generally easier to accommodate the observed Higgs mass than in weak scale SUSY. We will discuss the bounds from low energy experiments such as flavor physics in Section.\ref{sec:pheno}, where we can see that the cosmologically allowed regions are typically compatible with these constraints due to the large $m_0\gtrsim100$ TeV. \\
 
%\newpage
\section{Phenomenology, Signatures}\label{sec:pheno}
    \indent 
    
   The baryogenesis scenario proposed here has impacts on experimental signatures along several paths as follows. \\
   
 \underline{Low energy experiments}\\
 
%\textit{$B-(L-)$ violation:}\\

   By integrating out gaugino and sfermion masses the $\cancel{B}$ or $\cancel{L}$ operators essential for generating the asymmetry give rise to higher dimensional operators which may reveal themselves at low energy experiments. $\bar{u}_1\bar{d}_1\bar{d}_1$ type of operator can be effectively probed by $n-\bar{n}$ oscillation experiment. The current limit gives: $m_0\gtrsim10^6{\rm ~GeV}(g_X\lambda_{111}^{''})^{\frac{1}{2}}\left(\frac{1{\rm~TeV}}{m_X}\right)^{\frac{1}{4}}$\cite{Abe:2011ky}, where $m_X, g_X$ is the mass and coupling of the gaugino $X$ under consideration. Other flavor combinations of $\bar{u}\bar{d}\bar{d}$ couplings receive comparable or looser bound from other measurements such as $p\rightarrow K^+\nu$\cite{Goity:1994dq}. Meanwhile, the numerical results as discussed in Section.\ref{sec:results} favor $m_0\gtrsim10^5-10^6$ GeV (with $\lambda^{''}\sim O(0.1)$), therefore $\cancel{B}$ effect from our baryogenesis model can be well consistent with the current constraints while within reach of the next generation experiments \cite{Snow:2009zz, Balashov:2007zj}. The $LQ\bar{d}$ type of operator relevant to our leptogenesis model can be probed by experiments such as $\nu$-less double beta decay\cite{Babu:1995vh}, where the current bound is: $m_0\gtrsim10^4{\rm ~GeV}(g_X\lambda_{111}^{'})^{\frac{1}{2}}\left(\frac{1\rm~TeV}{m_X}\right)^{\frac{1}{4}}$\cite{Barbier:2004ez}. Constraints on other flavor combinations of $LQ\bar{d}$ operator can be found in \cite{Barbier:2004ez}. As can be seen from the results in Section.\ref{sec:results}, our leptogenesis model comfortably satisfies these limits and it may be reachable by future experiments\cite{Kaufman:2013ysa}. \\
   
%\textit{CP violation, Flavor violation:}\\

The $\cancel{CP}$ essential to the asymmetry generation and the flavor violating (FV) effect that may help promote the amount of the asymmetry can be experimentally tested as well. For instance, the flavor-violating operator of $dd\bar{s}\bar{s}$ type can be probed by $K_0-\bar{K}_0$ mixing, which currently provides constraint: $m_0\gtrsim10^4{\rm~GeV} ~Im(g_{X11}g^*_{X12})^{\frac{1}{4}}\left(\frac{m_X}{1\rm~TeV}\right)^{\frac{1}{2}}$, where $11, 12$ are the flavor indices for the couplings of the gaugino $X$ in the loop. The current limits from other related experiments such as neutron EDM, $\mu\rightarrow e\gamma$ on $m_0$ are at similar level, i.e., $m_0\gtrsim O(\rm TeV)$. The upgrades in the near future can probe the split SUSY scale up to $100-1000$ TeV\cite{Moroi:2013vya, McKeen:2013dma, Altmannshofer:2013lfa}, which would be sensitive to the parameter space of our models. \\

 \underline{LHC and its possible upgrades:} \\
 
   In addition to the above indirect searches from low energy experiments at the intensity frontier, high energy collider experiments provide opportunities to directly probe signals from our models. The baryon parent $\tb$, and the equally important lighter $\tilde{g}$ or $\tw$ which contributes to the interference loop, all lie around TeV scale or lower, so are within energy reach of the 14 TeV LHC or potential higher energy upgrade at 33 TeV or 100 TeV. Direct production of the baryon parent $\tb$ is pessimistic due to its highly decoupled nature. Secondary production of $\tb$ from cascade decay from a possibly heavier $\tilde{g}$ (in the leptogenesis model) or $\tw$ (in the baryogenesis model) can be promising. However, as shown in our analytic estimate around eqs.(\ref{epsiloncp_bg},\ref{epsiloncp_lg}) and numerical examples, typically a sizable asymmetry require $m_{\tb}\gtrsim$1 TeV, while the LHC reach for gluino mass is barely $\sim 2$ TeV, for wino mass $\sim$ 1 TeV\cite{Bhattacherjee:2012ed}. The production of $\tb$ in our model thus typically demands a higher energy upgrade of the LHC. Nonetheless, $\tilde{g}$ in our baryogenesis model or $\tw$ in the leptogenesis model is required to be lighter than $\tb$ in order to have an imaginary kinematic factor, and can be light enough to have a significant production rate at the LHC. After production, just like $\tb$, the light $\tg$ or $\tw$ dominantly undergoes 3-body decays via RPV couplings. The RPV decay rates of $\tg, \tw$ take the similar form to that of $\tb$ as given in eqs.(\ref{bv_decayrate},\ref{lv_decayrate}), which has a suppression factor of $(m_X/m_0)^4$. These decays also have $\cancel{B} (\cancel{L})$ and $\cancel{CP}$ features just like $\tb$. The cosmological late decay of $\tg$ or $\tw$ may contribute to baryon asymmetry as well, but by a small fraction. We can estimate the lifetime of $\tg,\tw$ using the analogy of eqs.(\ref{bv_decayrate},\ref{lv_decayrate}), and find that the decay length of $\tau_D\gtrsim1$ cm is typical for $m_0\sim100-1000$ TeV and $m_X\lesssim1$ TeV. Such macroscopic decay length can leave \textit{a displaced vertex} inside the detector, which is a general class of striking signal that has not been well explored in the conventional searches and drawn rising attention and endeavor recently\cite{Aad:2011zb, Graham:2012th,atlas_rpv,cms_dijet}. A challenging yet exciting further step is to \textit{measure the CP asymmetry} in the late RPV decays of the gauginos. Unlike the model presented in \cite{Cui:2012jh}, the CP asymmetry here is highly suppressed ($\epsilon\lesssim10^{-6}$) and hard to be measured at the LHC. A 33 TeV or 100 TeV collider would offer better opportunity to observe the asymmetry due to larger production rate and higher mass reach. We leave the detailed study on the collider phenomenology of our models to future work.

\section{Conclusions, Outlook}\label{sec:concl}

    In this work, we demonstrate that mini-split SUSY with RPV couplings can naturally provide all the ingredients for a successful baryogenesis mechanism, without any additional matter or structure beyond the minimal model (MSSM). The naturally late $\cancel{CP},\cancel{B}$ decay of bino after its thermal freezeout triggers baryogenesis. With gaugino masses $\sim O$ (TeV), cosmological conditions favor sfermion masses to be $\sim10^2-10^3$ TeV. It is rather intriguing that this happens to be around the ``mini-split'' scale which is independently motivated by the limits from flavor physics experiments as well as the higgs mass measurement. In order to get sufficient baryon asymmetry, $\mu\gg m_0$ is necessary in addition to the existing split of $m_0\gg m_{\rm gaugino}$. Since the $\mu$ term is the only supersymmetric parameter in the MSSM, despite related tuning, a large value as needed here, is as plausible and phenomenologically innocuous as the conventionally well considered case of $\mu\lesssim m_0$. The baryogenesis models presented here may inspire further studies on the UV explanation and implications of such a spectrum with large $\mu$. Furthermore, the criticality of a mini-split spectrum for a successful baryogenesis here suggests that the loss of full naturalness in SUSY may result from a compromise between naturalness principle and environmental selection. The results in this work therefore may be seen as an example analogous to the ``galactic principle'' or ``atomic principle'' stressed in earlier work such as \cite{ArkaniHamed:2004fb,Giudice:2006sn,Weinberg:1987dv,Agrawal:1997gf}.\\

    The potential interface between new cosmology and weak scale new particle physics such as those related to supersymmetry is a very attractive possibility based on both the theoretical motivations as well as the experimental testability. Inspired by the ``WIMP miracle'', the exploration of such connection has been mostly focused on the front of dark matter. This work, together with earlier recent work such as \cite{Cui:2012jh, Cui:2011ab,wimpbg}, where baryogenesis is demonstrated to be triggered by a WIMP type of particle, brings up a new perspective that the origin of baryon asymmetry $\Omega_{\Delta B}$--a cosmic phenomenology as important and puzzling as $\Omega_{\rm DM}$--can be another interface where new particle physics meets new cosmology. The search for signatures of these models can be multi-pronged, just as for the case of WIMP dark matter. At the energy frontier, i.e. collider experiments such as the LHC and its potential upgrades, a WIMP particle with cosmological late decay may reveal itself in the form of a displaced vertex. Furthermore, in some cases the CP asymmetry responsible for baryogenesis may be measurable at the collider experiments. At the intensity frontier, as discussed here and for the model presented in \cite{Cui:2012jh}, for a large model parameter space, the $\cancel{B}(\cancel{L}), \cancel{CP}$ and possible flavor-changing effects are within reaches of the near future upgrades of relevant low energy experiments. Detailed studies of these prospects would be very interesting. Hopefully the near future experiments at various frontiers may unveil even more mysteries than we expected, including the cosmic origin of our baryonic world...

\section*{Acknowledgements}
We thank C.~Cheung, F.~D'Eramo, S.~Martin, R.~Mohapatra, B.~Shuve and R.~Sundrum for discussion. We also thank the Kavli Institute for Theoretical Physics for hospitality while part of the work was conducted. This work was supported in part by NSF grant PHY-0968854 and by the Maryland Center for Fundamental Physics.

\end{document}